\documentclass[useAMS]{mn2e}

\def \XMM {$XMM$-$Newton$\ }

\usepackage{graphicx}
\usepackage{lscape}
\title[Orbital and Spin Phase-Resolved Spectroscopy of EX Hya Using XMM-Newton]
{Orbital and Spin Phase-Resolved Spectroscopy  of the Intermediate
Polar EX Hya Using \emph{XMM-Newton} Data}
\author[Pek\"on Y. and Balman, \c{S}.]{Y.
Pek\"on$^{}$\thanks{E-mail:yakup@astroa.physics.metu.edu.tr} and \c{S}.
Balman$^{}$\thanks{E-mail:solen@astroa.physics.metu.edu.tr} \\ \\
$^{}$Department of Physics, Middle East Technical University,
In\"on\"u Bulvar{\i}, Ankara, 06531, Turkey}

\begin{document}

\maketitle

\label{firstpage}

\begin{abstract}

We present for the first time orbital phase-resolved spectra of an
intermediate polar (IP), EX Hya, together with the spin
phase-resolved spectra during two different epochs using the X-ray
Multi-Mirror Mission (\XMM), European Photon Imaging Camera (pn
instrument). We find that the source at the two epochs has the same
X-ray luminosity of $\sim$ 6.5 $\times$ 10$^{31}$ erg s$^{-1}$. We
detect spectral variations between the 2000 and 2003 observations of
the source. We fitted the spectrum using a neutral hydrogen
absorption model with or without covering fraction together with
Gaussians for emission lines, two collisional equilibrium plasma
emission models (MEKAL) and a cooling-flow plasma emission model
(VMCFLOW). We find that two of the three emission components
($kT$=0.6-0.8 keV and $kT$=1.3-1.7 keV) fitted by the MEKAL models
are almost constant over the spin and orbital phases and also over
the two different epochs with the normalisation varying directly
proportional to the flux when the data are folded according to the
orbital and spin phase indicating that the slight variation may be
due to occultation. The emission modeled by the VMCFLOW changes over
the spin and orbital phases and the 2000 and 2003 observations
reveal two different ranges of temperatures (3-33 and 8-61 keV
respectively) that model the shock zone in the accretion column/s.
The ratios of the spin maximum to minimum and the orbital maximum to
minimum spectra along with the increase in the plasma temperatures
indicate that the spectrum gets harder in the minimum phases of both
orbital and spin periods.
%This indicates that ionized absorbers are present since
%absorption effects go up to 3-5 keV.
In the 2003 observation, a 6.4 keV fluorescent Fe emission line is
present at the orbital minima in a range of phases from 0.9 to 1.3
and it is absent otherwise. This indicates that there is reflection
from the disc most likely from a large bulge at the accretion impact
zone.

\end{abstract}


\section{Introduction}

 Cataclysmic variables (CVs) are compact binaries
hosting a white dwarf (WD) primary star accreting material from the
Hydrogen rich companion. The accretion occurs through an accretion
disc in cases where the magnetic field of the WD is weak ( $B$ $<$
0.01 MG) and such systems are referred to as nonmagnetic CVs. The
magnetic CVs (mCVs) constitute about 25$\%$ of the entire population
of CVs. There are two subclasses depending on magnetic field
strength. Polars (AM Her type systems) have $B$ $>$ 10 MG where this
strong field causes the accretion flow to directly channel onto the
magnetic pole of the WD inhibiting the creation of an accretion disk
(see Cropper 1990; Warner  2003). Polars show strong circular and
linear polarization modulated at P$_{\rm{orb}}$ and the WD rotation
P$_{\rm{spin}}$ is synchronized with the binary orbit. Only about
1$\%$ of the Polars are found to show asynchronicity. The second
class of mCVs are Intermediate Polars (DQ Her type systems) that
have a weaker field strength compared with Polars of about 1-10 MG.
In this case, the accretion takes place via a truncated disk to the
magnetic poles through accretion curtains. Near the surface of the
WD, the accreting material forms a strong shock where the post shock
region heats up to 10-20 keV and then cools via thermal
Bremsstrahlung (Patterson 1994; Hellier 1996; de Martino et al.
2008; Brunschweiger et al. 2009). IPs are asynchronous systems where
the orbital period of the system P$_{\rm{orb}}$ is larger than the
spin period of the white dwarf P$_{\rm{spin}}$. The majority of the
IPs have P$_{\rm{spin}}$/P$_{\rm{orb}}$ $<$ 0.1 , theoretically they
are predicted to lie in the 0.01-0.6 range (Norton, Wynn \&
Somerscales 2004; Norton et al. 2008; Scaringi et al. 2010). IPs are
well characterised by high complex absorption in their X-ray spectra
which extends, where the spectra are available in some cases, out to
100 keV. They show modulation of the X-ray and/or optical light
curves at P$_{\rm{spin}}$ and the appearance of the beat period
(where 1/P$_{\rm{beat}}$ = 1/P$_{\rm{spin}}$ - 1/P$_{\rm{orb}}$) or
its sidebands. Many  mCVs are detected by $INTEGRAL\ IBIS$ (Barlow
et al. 2006; Landi et al. 2009), and $Swift\ BAT$ (Brunschweiger et
al. 2009) in the 20-100 keV ranges and most of them are found to be
IPs. This is thought to be because the hardest X-ray emission comes
from the cooling post-shock region of IPs rather than Polars, since
the strong magnetic field of Polars causes the cyclotron cooling to
dominate the cooling post-shock region suppressing the
Bremsstrahlung emission.

EX Hya was discovered by Kraft (1962) with an orbital period of 98
minutes. Later, the 67 minutes rotation period of the white dwarf
was identified (Vogt et al. 1980), classifying the source as an
intermediate polar. The WD mass is estimated to be in the range of
0.5-0.8 M$_{\odot}$ as measured using optical (Beauerman et al.
2003; Hellier et. al 1987) and X-ray (Fujimoto \& Ishida 1997)
observations. The elemental abundances are in accordance with solar
abundances except for iron which is less than solar (0.6 $\pm$ 0.2)
(Fujimoto \& Ishida 1997). Using optical observations, the system
has been found to be at a distance of 64.5 $\pm$ 1.2 pc, with a
quiescent accretion luminosity of (2.6 $\pm$ 0.6)  $\times$
10$^{32}$ erg s$^{-1}$, a quiescent accretion rate of (6.2 $\pm$
1.5) $\times$ 10$^{-11}$ M$_{\odot}$ yr$^{-1}$ and an inclination of
76.0$^\circ$ - 77.6$^\circ$ (Beuermann et al. 2003).

EX Hya shows a partial eclipse in its light curves over the orbital
period. The eclipse was explained by two scenarios; either
occultation of one of the poles for two-pole accretion or partial
occultation of the accretion column by the secondary for one-pole
accretion (Rosen, Mason \& Cordova, 1988). The source is suggested
to have an extended bulge on the outer accretion disc and
extended/overflowing material originating from the hot spot (Belle
et al. 2005). The source is observed both in outburst and
quiescence. During the outburst stage, the Doppler tomograms and
radial velocities indicate that the source shows disc overflow and
dips in the optical light curve (Mhlahlo et al. 2007a); the disc
overflow is also present in the quiescent state (Mhlahlo et al.
2007b) consistent with the above scenario. The evidence of a bulge
on the accretion disc is also present in the X-ray observations. The
X-ray binary light curves in the 1-20 {\AA} range show an extended
broad absorption associated with the disc which increases as the
wavelength increases (Hoogerwerf, Brickhouse \& Mauche 2005).

The X-ray spectrum of the post shock emission of EX Hya has been
previously modelled with multi-temperature plasma emission models.
One example uses four optically thin plasma emission models at four
different temperatures (Allan, Hellier \& Beardmore, 1998). Another
approach uses a cooling- flow model (MKCFLOW) assuming a
multi-temperature thermal plasma emission with a relatively flat
emission measure distribution indicating cooling gas from a
steady-state condition. (Mukai et al. 2003).

The spin modulation of the X-ray light curve of the source can be
explained either by the absorption difference over the spin phase
because of the accretion curtain, or occultation of the accretion
curtain by the WD. The occultation scenario is more favoured because
the system has tall accretion coloumns ($\sim$ 1 R$_{\rm{WD}}$ )
(Allan, Hellier \& Beardmore, 1998). The source also shows spectral
variability as a function of orbital phase due to absorption
(Cropper et. al 2002), but the modulations and spectral models have
not been thoroughly investigated up until this work.

In this work, the X-ray spectral properties of EX Hya are thoroughly
investigated. In Section 2, the X-ray observations and data
preparation are introduced. In Section 3 the total spectrum for the
2000 and 2003 data are investigated. A composite-model fit for both
observations is performed and the variation of the spectral
parameters between the two observations are discussed. In Section 4,
the characteristics of the source spectra in both 2000 and 2003
observations are studied over the orbital phase. The trends in the
derived spectral parameters as a function of the orbital phase as
well as the differences between the 2000 and 2003 data are compared.
In Section 5, a similar analysis is made, this time as a function of
the spin phase of EX Hya. In Section 6, the results are discussed
and finally in Section 7 the summary and conclusion of our work is
presented.

\section{Observation and Data}

Two different archival observations of EX Hya obtained with the \XMM
Observatory were used in this study.  The first one was conducted on
01 July 2000 (OBS ID: 00111020101) with 30 ksec exposure time and
the second one on 11 January 2003 (OBS ID:0057740301) with 57 ksec
exposure time. The \XMM Observatory (Jansen et al. 2001) has a pn
CCD detector (Str\"{u}der et al., 2001) and two MOS CCD detectors
sensitive in the 0.2-15 keV energy range at the focus of three
European Photon Imaging Cameras (EPIC) with a field of view of about
half a degree for each. It also has Reflecting Grating Spectrometers
(RGS) (den Herder et al. 2001), a high resolution spectrometer
working together with the EPIC detectors sensitive in the 0.35-2.5
keV energy range as well as a 30 cm optical monitor instrument (OM)
with an optical/UV telescope (Mason et al. 2001).

We first checked the pile up in the observations using {\sc
epatplot}\footnote{\scriptsize See
http://xmm.esac.esa.int/sas/current/documentation/threads/epatplot.shtml}
tool in {\sc sas}. The EPIC MOS data were piled up in the 2000
observation, hence only EPIC pn data was used for both observations
in this analysis. The 2000 observation was a pointed observation of
the source in the small window mode for the EPIC pn with a net count
rate of 40.1 $\pm$ 0.04 count s$^{-1}$ in the 0.2-10 keV energy
interval. In the 2003 observation the EPIC pn was used in the
extended full frame mode and the source is off axis by 14
arc-minutes, the net source count rate is 9.7 $\pm$ 0.01 count
s$^{-1}$ in the 0.2-10 keV energy range. The significantly lower
count rate in the 2003 observation is due to vignetting, caused by
the source being off-axis. We utilized the {\sc calview} tool in
{\sc sas} and determined the percentage of loss in the count rate as
65$\%$ which yields a count rate of about 13-14 count s$^{-1}$ based
on 40.1 count s$^{-1}$ from 2000. This is in accordance with our
count rate derived from the 2003 data. However, we note a difference
of about 30-40$\%$ between the rate calculated from the data and our
inspection from {\sc calview} using the rate of the 2000 data. This
may be a result of extra vignetting (corrected in the analysis) or
about 30-40$\%$ change in the source count rate, which is not
conclusive at this stage.

For both observations, the standard pipeline processed data was used
to perform the spectral and temporal analysis using the \XMM Science
Analysis Software ({\sc sas}) version 8.0.0. The source light curve
and spectrum extraction are done using the {\sc xmmselect} tool
within {\sc sas}. In the 2000 observation, a circular photon
extraction region was centered on the source with a radius 38.8
arcsec, and a same size region was used for the background
extraction centered elsewhere on the image. For the 2003 data, since
the source was off-axis, an elliptical region with a semi-major axis
of 60 arcsec and a semi-minor axis of 23.3 arcsec, orientated at an
angle of 320 degrees was used in order to extract the source and
background events. Data with single- and double-pixel events, i.e.,
patterns 0--4 with Flag=0 options were used. In order to perform
spectroscopy, we used the {\sc phasecalc} tool in {\sc sas}, and
created phase columns in the event files using the spin and orbital
periods; and spectra were extracted from the event files for a phase
interval of 0.1  with iterative calls to the {\sc especget} tool in
{\sc sas} yielding phase-resolved spectra. {\sc xspec} 12.2.1
(Arnaud 1996) was used for further analysis.
%All the  spectra were grouped to have
%a minimum of 100 counts per spectral bin in order to improve the
%statistical quality for the subsequent fitting procedure.

\section{Total Spectrum of the Source}

Prior to the fitting procedure, spectra were calculated using the
extraction criteria described in section 2 in the energy range of
0.2-10.0 keV. The resulting spectra were grouped so that the minimum
counts for each energy bin were 300 for the year 2000 observation
and 200 for the 2003 observation using {\sc xspec}. The fits were
conducted in the 0.2-10.0 keV energy band.

The first attempts at fitting the total spectra using {\sc xspec}
were done by using a plasma emission model that is in collisional
equilibrium (MEKAL model in {\sc xspec}) and creating a composite
model of several MEKAL models to fit the spectra at several
different temperatures as suggested by Allan et al. (1998). Although
we included up to four different MEKAL model components, the trials
yielded fits with reduced $\chi^2$ more than 2. Then a different
approach was taken by using a cooing-flow model (the VMCFLOW model
in {\sc xspec}) suggested by Mukai et al. (2003) which again failed
to yield good fits. Finally, we combined the two methods, using a
cooling-flow model (VMCFLOW) together with two different temperature
MEKAL models at around 0.64 keV and 1.6 keV. An absorber model for
the interstellar extinction and a partially covering absorber model
(TBABS and PCFABS models in {\sc xspec}) was assumed to account for
the intervening absorption for both of the observations.

The fits with the multi-component model yielded fluctuating
residuals around particularly the iron line complex at about 6.7
keV. We believe this problem arises from CTI effects around
particularly strong lines in the EPIC pn spectrum. Thus, we decided
to model the lines separately and  decreased the Fe abundance in the
VMCFLOW model to 0.2 times the solar abundance and added Gaussian
emission lines to the model at 6.7 keV and 6.9 keV to account for
the Fe$\alpha$ emission line complex. This procedure enabled us to
individually model the Fe emission lines. We investigated any
effects of the reduction of the Fe abundance in modeling the spectra
and we detected no other effects. This is expected since the VMCFLOW
models only the harder part of the spectra. Other Gaussian emission
lines were included in the fitted composite model at 7.8 keV, 0.58
keV and 0.78 keV for the 2000 data and 7.8 keV, 6.4 keV and 0.58 keV
for the 2003 data in order to reduce the high sigma deviations in
the residuals. The 7.8 keV, 0.58 keV and 0.78 keV lines correspond
to Fe XXV K$\beta$, O VII and O VIII transitions respectively
according to the CHIANTI atomic database. Modeling these lines by
increasing the related element abundances is not possible, hence
Gaussian profiles are used instead. The fitted spectra are presented
in Figure 1 and the spectral parameters are displayed in Table 1.

We calculate unabsorbed fluxes with similar values of 1.43 $\times$
10$^{-10}$ and 1.24 $\times$ 10$^{-10}$ erg cm$^{-2}$ s$^{-1}$ for
2000 and 2003 observations, respectively. These flux values yield
luminosities of 6.7 $\times$ 10$^{31}$ erg s$^{-1}$ and 5.8 $\times$
10$^{31}$ erg s$^{-1}$ assuming a 64.5 pc source distance. Using the
relation L=G$\rm{\dot{M}}$M/2R and taking the mass of the white
dwarf as 0.5 M$_{\odot}$ and radius as 0.95 R$_{\odot}$ (Beuermann
et al. 2003), we find mass accretion rates of 5.75 $\times$
10$^{-11}$ M$_{\odot}$ yr$^{-1}$ and 5.0 $\times$ 10$^{-11}$
M$_{\odot}$ yr$^{-1}$ for 2000 and 2003 data respectively.

Both spectra show an unusual feature; an emission line at around 7.8
keV. To investigate this feature more closely, the spectra of the
2000 and 2003 data, were fitted over a range of 5-10 keV, with a
power law model for the continuum and three Gaussians to model the
emission lines at around 6.7 keV, 6.9 keV and 7.8 keV. The resulting
parameters of the continuum and the spectral lines are displayed in
Table 2. We find an emission line centered around
7.8$^{+0.3}_{-0.3}$ keV with a line width $\sigma$ =
0.06$^{+0.06}_{-0.06}$ keV for 2000, and centered around
7.9$^{+0.8}_{-0.8}$ with a line width $\sigma$ = 0.5$^{+0.3}_{-0.1}$
for 2003 observations. We believe this is the Fe XXV K$\beta$
emission line. The line widths for the 6.7 keV emission line is
similar for both observations. In the 2003 data, the line widths for
6.9 and 7.8 keV lines are significantly higher.

\section{Orbital Phase-Resolved Spectroscopy}

In this work, a detailed orbital phase-resolved spectroscopy of EX
Hya was carried out for the first time. For both observations, 2000
and 2003, phase columns were created using the ephemeris T$_0$ =
2437699.94179+0.068233846(4)E given by Hellier \& Sproats (1992).
Then for each 0.1 phase interval from phase 0 to 1, the spectra were
extracted as described in section 2. The calculated spectra were
grouped such that each energy bin contained minimum of 100 counts to
improve the statistical quality. The channels below 0.3 and above
10.0 keV were ignored during the fitting procedure to improve the
statistical quality of the spectra.

To start with, we calculated the ratio of the maximum to minimum
spectrum of the orbital phase-resolved spectra in order to study the
effect of absorption/occultation in the system in the two different
years. The ratio of the normalised count rate per energy for phases
around the orbital maximum (phases between 0.6-0.8) to phases around
the orbital minimum (phases between 0.9-1.1) are plotted against the
energy in Figure 2. Both 2000 and 2003 spectral ratios show clear
spectral variation and a power-law decline as the energy increases
 with ratio values greater than 1
particularly at lower energies. There is a more pronounced steep decline with
increasing energy in the year 2003 showing the existence of large
scale absorption difference within the system from orbital maximum to minimum.
%that shows since the ratios show clear persistence over 1
%even at high energies (up to 3 keV). For the 2000 data, the ratios
%decline slowly from about 3 to 2 with the increasing energy. In 2003
%data, the ratios are as high as 5 and show a steeper decline down to
%1 as the energy increases (see Figure 2).

As in the case of the fits to the total spectra, models such as
single MEKAL (in {\sc xspec}), composite multiple MEKAL or a single
VMCFLOW models alone yielded large reduced $\chi^2$ values above 2
during the fitting procedure within {\sc xspec}. Thus, we attempted
to fit a combination of a MEKAL model with a VMCFLOW  using solar
abundances along with a photoelectric absorption model (WABS in {\sc
xspec}). However, this approach did not yield good fits either. The
Fe emission line complex around 6.7 keV still yielded fluctuating
residuals. Thus, we applied the same procedure to reduce the Fe
abundance and adding individual Gaussian lines to model the 6.7 keV
and 6.9 keV iron lines when necessary.

In the analysis of the 2000 data, we used a  composite model to fit
the spectra at different phases comprising of two collisional
equilibrium plasma emission models, MEKAL1 and MEKAL2, one
cooling-flow plasma emission model, VMCFLOW, a photoelectric
absorption model, WABS and three Gaussians for particular Fe lines.
The spectral parameters from the fit results are displayed in Table
3 and selected parameters are plotted against the orbital phase in
Figure 3. The fitted spectra at orbital minimum (phase 0.9) and
maximum (phase 0.3) are shown in Figure 4. For the absorption
component, a simple cold absorber was enough to fit the data, hence
a partial covering absorber was not used. The column density of
neutral Hydrogen (absorption parameter ($N_{\rm H}$)) shows distinct
variation with the orbital phase and indicates an anti-correlation
with the orbital flux variations reaching a maximum at the orbital
minimum. This clearly shows that orbital variations at X-ray
wavelengths are produced by absorption on the orbital plane. The
$LowT$ (low temperature) parameter of the VMCFLOW model shows the
same anti-correlation with flux and reaches a minimum at orbital
maximum whereas the $HighT$ (high temperature) parameter of the same
model shows almost a direct correlation with the flux which may
imply scattering being at work. The plasma emission components
modeled by the MEKAL1 and MEKAL2 models show almost a constant
temperature of about 0.6-0.8 keV and 1.3-1.7 keV and the
normalisation varies slightly in phase with the orbital motion. The
Gaussian line at 6.7 keV does not vary with orbital phase apart from
a slight drop between phases 0.6-0.7 which indicates that the region
where the line is produced is visible at all times during the binary
motion.

%As in 2003 data, the nH values are higher at minimum phases and low at maximum phases.
%Here, 6.7 and 6.9 keV emission lines were present at all phases, on
%the other hand 6.4 keV lines were absent at all phases. Both for
%2000 and 2003 data, the spectral parameters for each phase are
%plotted against the phase to demonstrate the change of physical
%parameters during the orbital motion (Figure 3).

In the analysis of the 2003 data, we used a  composite model
comprising two collisional equilibrium plasma emission model, MEKAL1
and MEKAL2, one cooling-flow plasma emission model, VMCFLOW, a
photoelectric absorption model, WABS and three Gaussians for
particular Fe lines. The spectral parameters from the fit results
are displayed in Table 4 and selected parameters are plotted against
the orbital phase in Figure 5. The fitted spectra at orbital minimum
(phase 0) and maximum (phase 0.4) are shown in Figure 6. The
resulting fits have reduced $\chi^2$ well below 2 for all the phase
spectra. We find that some parameters show distinct variations with
the orbital phase. The $N_{\rm H}$ parameter for the intrinsic
absorption and $LowT$ parameter of the VMCFLOW show anti-correlation
with the flux i.e. they yield higher values at orbital minimum
phases (phases between 0.9 and 1.3), and lower values at the maximum
phases (phases between 0.4 and 0.8). This indicates that source
emission is being absorbed at particular phases (0.9 and 1.3) where
there is absorption on the orbital plane and the spectrum gets
harder in general (softer X-rays are absorbed). The $HighT$
parameter shows no distinct variation over the orbital phase in the
2003 data. This is obvious since harder X-rays will not be
affected/absorbed as much by the neutral absorption.

The plasma emission components modeled by the MEKAL model in the
2003 data show almost a constant temperature of about 0.6-0.8 keV
and 1.3-1.7 keV; the normalisations vary slightly in phase with the
flux. These components may be affected by only occultation , i.e.,
the eclipse itself. The width $\sigma$ and normalisation of the 6.7
keV emission line shows direct correlation, with low values at
orbital minimum and high values at orbital maximum. The 6.7 keV
emission line (Fe XXV line) is present at all phases. For the
orbital minimum (phases between 0.9 and 1.3), emission lines at 6.9
keV (Fe XXVI line) and at 6.4 keV (fluorescent Fe line) were
necessary for successful fits to the data (See Figure 6).  Figure 7
shows the necessity to include the 6.4 and 6.9 keV emission lines in
the 2003 orbital spectra at minimum phases by fitting 4 MEKAL models
with variable abundances and focusing on the 6-8 keV regime. The
excess around 6.4 keV is clear in Figure 7. For the orbital maximum
(phases between 0.4 and 0.8), the 6.9 and 6.4 keV emission lines
could not be fitted with free parameters, however, upper limits for
normalisations were obtained by fixing all the other model
parameters.

The 6.7 keV line widths show no distinct variations in the 2000
data, with values around 0.1 keV. On the other hand, in the 2003
data, the widths vary in phase with the flux; the line is broader at
the phases of orbital maximum with widths around 0.3 keV, and at
phases of orbital minimum it has similar values to the 2000 data,
around 0.1 keV. The width of the 6.4 keV line in the 2003 data
varies between 0 and 0.16 keV.

\section{Spin Phase-Resolved Spectroscopy}

We have also performed spin phase-resolved spectroscopy in a similar
fashion to the orbital phase-resolved spectroscopy for the 2000 and
2003 data. The ephemeris for the spin period  T$_0$ =
2437699.8914(5) + 0.046546504(9)E given by Hellier \& Sproats (1992)
is used for both observations to create spin phase-resolved spectra
using the method described in section 2. {\sc xspec} was used for
the remaining spectral analysis, mainly the fitting procedure. The
spin phase-resolved spectra were grouped so that each energy bin
contained a minimum of 100 counts and data points below 0.3 keV and
above 10.0 keV were omitted due to bad statistics.

The ratio of the spin maximum spectra (phases between 0.75 and 0.95)
to spin minimum spectra (phases between 0.25 and 0.45) was also
studied for the spin phase-resolved spectra. For both the 2000 and
2003 data, the maximum to minimum ratios show similar behaviour,
thus there is no difference in the spin phase-resolved spectra, as
far as the absorption characteristics are concerned between the two
different observations of EX Hya. At low energies, the ratios have
values of about 2 that gradually decrease with increasing energy
down to 1 (see Figure 8).

Similar to the previous fits to spectra, two MEKAL models, a VMCFLOW
model and various Gaussians to model emission lines were used to fit
all the spectra at all spin phases (with 0.1 phase increments).
Differently from the orbital phase-resolved spectroscopy, inclusion
of a partial covering absorber model (PCFABS in {\sc xspec}) was
necessary to achieve good fits in the low energies.

The 2003 spin phase-resolved spectra were fitted with {\sc xspec}
models, two MEKAL  (collisional equilibrium plasma emission model
with solar abundances) at different temperatures, VMCFLOW
(cooling-flow model), PCFABS and a Gaussian for the emission line at
6.7 keV. No other emission lines were needed to be included in the
fit. Figure 9 shows the fitted spectra at the spin phase maximum
(phase 0.7) and minimum (phases 0.3). Table 5 displays spectral
parameters and Figure 10 is the plot of selected spectral parameters
versus the spin phase for the 2003 data. The lower and upper
temperature of the cooling flow (i.e. $LowT$ and $HighT$ parameters
of the VMCFLOW model) show anti-correlation with the light curve
folded over the spin period. The $N_{\rm H}$ and the covering
fraction of the partial covering absorber parameter does not show a
direct anti-correlation with the flux as would be expected from
neutral absorption producing the X-ray modulations. The spectral
parameters of the 6.7 keV line do not show any variation over the
spin phase. As with the orbital phase-resolved spectroscopy, the
need for emission lines were investigated by fitting the data with 4
MEKAL models with varying abundances and close inspection over the
6-8 keV range. The excess at 6.4 keV does not show any significant
pattern as in the orbital phase spectra. The 6.4 keV and 6.9 keV
emission lines could not be fitted with free parameters. However the
upper limits for the 6.4 and 6.9 keV line normalisations could be
calculated, with values 1.0 $\times$ 10$^{-8}$ and 5.2 $\times$
10$^{-7}$ for 6.9 keV and 2.7 $\times$ 10$^{-6}$ 1.4 $\times$
10$^{-6}$ for 6.4 keV at minimum flux (phases 0.2 and 0.3 ); 9.4
$\times$ 10$^{-10}$ and 8.5 $\times$ 10$^{-6}$ for 6.9 keV and 2.3
$\times$ 10$^{-6}$ and 6.0 $\times$ 10$^{-7}$ for 6.4 keV at maximum
flux (phases 0.7 and 0.8). This may be consistent with our results
since we find components and emission lines (like the 6.7 keV line)
that show widths and normalisations independent of phase generally
except for orbital phase in 2003.

The fits to the spin phase-resolved spectra of the 2000 data show
that the 6.7 keV and 6.9 keV emission lines were persistent at all
phases. In 2000 data,  the parameters $N_{\rm H}$, covering
fraction, $LowT$ and $HighT$ have higher values during orbital
minimum (phases between 0.1 and 0.4) and lower values otherwise
showing anti-correlation with the flux. This is the generally
expected case if absorption is at work creating the modulations and
the spectrum gets harder towards the minimum as the softer photons
are absorbed. The spectral parameters of the 6.7 keV line do not
show any variation over the spin phase in the 2000 data. The 6.9 keV
line is also persistent over the entire range of spin phase except
that the normalisation minimizes around phases 0.4-0.5 which is
about the spin minimum. Figure 11 shows the fitted spectra at the
spin phase maximum (phase 0.9) and minimum (phase 0.5). Table 6
presents the spectral parameters and Figure 12 displays the plot of
selected spectral parameters versus the spin phase.

The MEKAL emission components do not show any significant variation
of the temperature parameter over all phases, but a slight variation
of the normalisation in phase with the spin modulation for both 2000
and 2003 observations, is seen.

The widths and normalisations of the emission lines show no
significant variation with spin phase for both 2000 and 2003 data,
while the width of the 6.7 keV line in 2003 data is slightly larger
than that of 2000.

\section{Discussion}

We have presented the  \XMM EPIC pn data of the intermediate polar
EX Hya. The spectrum of EX Hya is quite complex as it needs
multi-component emission models to fit the data. For the total
spectra of both 2000 and 2003 observations, the composite model is
composed of 2 different MEKAL models with plasma temperatures around
0.6-0.8 keV and 1.3-1.7 keV, a cooling-flow emission model, VMCFLOW
with a distribution of 3-33 keV for the 2000 observation and 8-61
keV for the 2003 observation, and various Gaussian emission lines at
6.4 keV, 6.7 keV, 6.9 keV, 0.58 keV, 0.78 keV and 7.8 keV. The 7.8
keV emission line is an unusual feature encountered in the X-ray
spectrum of CVs. After a closer investigation around the 5-10 keV
spectra of both observations, the line appeared prominently with a
centre around 7.8$^{+0.3}_{-0.3}$, line width of 0.2-0.4 keV and
with a normalisation of 0.0001 photons cm$^{-2}$ s$^{-1}$. This line
may be the Fe XXV K$\beta$ emission line according to the CHIANTI
atomic database. The emission lines at 0.58 and 0.78 keV correspond
to O VII and O VIII transitions respectively. These lines appear
only in the phase averaged spectra, hence may be calibration
artifacts.

The two different collisional equilibrium plasma emission components
around 0.6-0.8 and 1.3-1.7 keV  (modeled with MEKAL) are present at
all phases (orbital and spin) and in the total spectra. This shows
that there is an X-ray emitting region unaffected by orbital and
spin phase which may be attributed to an X-ray emission region in
the line of sight at all phases. We find that the normalisations of
the MEKAL emission components decrease during the eclipse, which
indicates that these components originate most likely from the
post-shock region in the accretion column which is too complex to be
represented with just a cooling-flow plasma emission model alone.

The folded light curve of the source over the orbital period shows
two distinct features for both 2000 and 2003 data. The first feature
is the eclipse expected to be seen at phases around 0. For the 2000
light curve the eclipse falls at phases 0.93 $\pm$ 0.05. Fitting a
simple sinusoid to the folded light curve yields a semi-amplitude
variation of 6.4 $\pm$ 1.0 counts s$^{-1}$ and 14\% modulation. For
2003 data the eclipse is at phase 1.02 $\pm$ 0.05 with
semi-amplitude variation of 1.6 $\pm$ 0.3 counts s$^{-1}$ and 20\%
modulation. The accumulated errors in orbital phase are 0.000083 for
2000 and 0.000088 for 2003 data. With these values, it is clear that
the two eclipses do not overlap, however early works (e.g.
Hoogerwerf, Brickhouse \& Mauche 2005) show that the centroid of the
eclipse at high energies (i.e. for 10-15 {\AA} it is at phase 0.992)
falls behind the eclipse at low energies (i.e. for 15-20 {\AA} it is
at phase 1.0). Furthermore EUV eclipse was found to be at phase 0.97
by Belle et al. (2002) and it was suggested that the optical eclipse
at phase 0 might be due to the eclipse of the hot spot rather than
the white dwarf. Hence, the shifting of the eclipse may be to the
change in geometry of the eclipsing region between two observations.
During the eclipse, the count rate does not drop to zero, which is
consistent with a double pole accretion model where the lower
accretion column is eclipsed but the upper column is not (Belle et
al. 2005). This is consistent with our results since we find
components and emission lines (like the 6.7 keV line) that show
widths and normalisations independent of any type of phase. However,
it is inconclusive since the spin-folded light curves show a single
peak profile at all times.

The second prominent feature of the folded mean light curve over the
orbital phase is the broad shallow dip appearing in both 2000 and
2003 data. In the 2000 data, the dip is at around phases 0.45-0.65
and roughly 1/3 of the depth of the eclipse. In the 2003 data, the
absorption feature is shallower, (about 1/5 of the depth of the
eclipse) and the location of the dip is around phases 0.35-0.55.
This broad modulation was well observed over the optical, UV and
X-ray regime and attributed to the bulge and/or hot spot on the disc
(e.g. Belle et al. 2005, Mhlahlo et al. 2007, Hoogerwerf et al.
2005). The dip gets stronger and moves to phases closer to the
eclipse as the wavelength increases due to the changes in the
ionization state of the bulge. In the \emph{RXTE} light curve, the
dip is around phase 0.65 (Belle et al. 2005) , and in \emph{Chandra}
light curve it is around phase 0.78 (Hoogerwerf et al. 2005). Both
of these observations are from May 2000, and the broad dip feature
in the July 2000 \XMM light curve is in accordance with the above
examples.
 However the feature in 2003 is not consistent. The dip is
shallower and at phases further away from the eclipse. This may
imply that the source has undergone structural changes between 2000
and 2003 and the bulge has shifted position or grown in size so that
modulation is observed at earlier phases. Moreover, the
phase-resolved spectra show no significant change in the absorption
for either 2000 or 2003 observations, hence the modulation in the
orbital light curve cannot be attributed to neutral absorption on
the disc.

In both 2000 and 2003 data, the neutral absorption and low
temperature values of the cooling flow increase significantly during
the eclipse. This is an indication that cold absorption plays an
important role in the orbital variation of the source. During the
eclipse, the absorption increases, reducing the emission from the
lower end of the flow and hence the lower temperature values
increase. The change in absorption over the orbital motion has been
encountered before in another CV; a classical nova RR Pic (Pek\"{o}n
\& Balman 2008) where an absorbing region on the disk in the line of
sight affects the spectrum. This makes the spectral behaviour of CVs
over the orbital period an area worth more investigation.

The light curve folded over the spin period of the white dwarf shows
clear sinusoidal modulation for both 2000 and 2003 data. A simple
sinusoidal fit shows that the spin minimum is at phase 0.26 $\pm$
0.03, the semi-amplitude variation is 2.1 $\pm$ 0.1 counts s$^{-1}$
with 25\% modulation for 2003 data. For the 2000 data the minimum is
at phase 0.36 $\pm$ 0.06, the semi-amplitude variation is 14.5 $\pm$
1.2 counts s$^{-1}$ with 30\% modulation. The accumulated errors in
phase are 0.00027 for 2000 and 0.00029 for 2003 data, thus they are
much smaller than our phase errors. There is a difference in the
phase of the spin minimum in different years, most likely due to the
changes of the accretion column/curtain structure.

The fits show a net increase of absorption coloumn and covering
fraction at the spin minima in the 2000 data indicating the role of
absorption from the curtains. Moreover, the low temperature limits
of the plasma increase at the phases of minimum flux, making the
spectrum harder; since absorption reduces the amount of low
temperature X-rays. In the 2003 data the absorption coloumn and
covering fraction do not show a clear correlation with the flux, but
the low temperature limits of the plasma modulate inversely with the
flux. These findings imply that the role of absorption from the
accretion curtain is dominant for the source. Although our results
do not rule out occultation by the white dwarf as proposed by Allan
et al. (1998), the findings favour the absorption effects on the
spin modulation to be more dominant.

X-ray spectra of IPs commonly show the fluorescent 6.4 keV emission
line feature originating from the reprocessing of the hard X-rays
from the white dwarf surface (e.g. Hellier \& Mukai 2004, Landi et
al. 2008). The 6.4 keV line was encountered in the \emph{Chandra
HEG} spectrum of EX Hya (Hellier \& Mukai 2004), where the line was
weaker than the other lines in the Fe K$_{\rm{\alpha}}$ complex
(i.e. 6.7 and 6.9 keV emission lines). The total spectrum of the
2003 \XMM observations agrees with this, since the normalisation of
the Gaussian fitted to the 6.4 keV line is less than those of the
6.7 and 6.9 keV emission lines. The behaviour of this fluorescent
line over the orbital cycle is however unusual. In the
phase-resolved spectra, the line is detected for the orbital and
spin phase-resolved spectra of the 2000 observation and the spin
phase-resolved spectra of the 2003 data. In the analysis of the 2003
data over the orbital phases, the line is undetected at phases 0.4
to 0.8 (i.e., phases of orbital maxima) but present otherwise (i.e.,
phases of orbital minima). This means that there is a region on the
orbital plane reflecting the X-ray emission seen around the eclipse
phases which is a region on the disc rather than the surface of the
white dwarf. This reflection phenomenon is detected from a CV and an
IP for the first time. We strongly suggest that the bulge at the
accretion impact zone will appear at similar binary phases to the
eclipse and causes this reflection to occur. Such features are
detected in dipping Low-mass X-ray binaries (Diaz-Trigo et al.
2006), although with significantly higher line widths (e.g. $\sigma$
$\sim$ 0.85 keV) than seen in this work ($\sigma$ $\sim$ 0.1 keV).

It has been shown that a growing number of IPs have a soft blackbody
emission component with temperatures less than 100 eV, (e.g. Evans
\& Hellier 2007, Anzolin et al. 2008, Anzolin et al. 2009, de
Martino et al. 2008). EX Hya is listed as a "soft-IP" by Evans \&
Hellier (2007) however not with much emphasis, since the spectral
fits with a blackbody component yield poor results. Our fits with a
blackbody component also  give reduced $\chi^2$ values much greater
than 2, thus we are also unable to list the source as a "soft-IP".
There is no need for a soft blackbody component to model the \XMM
data of EX Hya for both observations.

The cooling flow-plasma emission component yield temperature ranges
3-33 keV (for 2000 data) and 8-61 keV (for 2003 data) in accordance
with the high plasma temperatures obtained from other IPs as
expected (e.g. Anzolin et al. 2008, Anzolin et al. 2009, de Martino
et al. 2008). We note that given these observations (e.g.,
Brunschweiger et al. 2009) EX Hya is one of the IPs that show the
hardest X-ray emission. In previous works with different
observations the spectra showed different temperature distributions.
Such measurements include 19.4 keV from \emph{SWIFT} data
(Brunschweiger et al. 2009), 0.54-15.4 keV from \emph{ASCA} data
(Allan et al. 1998), $<$ 20 keV from \emph{Chandra} data (Mukai et
al. 2003) and were obtained by fitting plasma models to the spectra,
extending up to $\sim$80 keV (\emph{Chandra HETG} data) using
measurements of the broad components of emission lines form the
pre-shock flow (Luna et al. 2010).

The count rate ratio plots of the spin maximum to minimum and
orbital maximum to minimum gives us an idea about the absorber
characteristics. The 2000 and 2003 spin ratios are similar, starting
at a value about 2 and gradually falling down to 1 with increasing
energy. This may imply that the absorption due to the accretion
curtain has similar structure in both 2000 and 2003. However, we
find that the neutral hydrogen column density is about a factor of
four larger in the 2000 data compared with that in 2003. So the
absorption in the accretion column/curtain is removed to some extent
in the 2003 data. The orbital spectral ratios show a significant
diffence in the years 2000 and 2003. They both show a power-law
decay over the \XMM\ energy range, where the 2000 data is flatter
and the 2003 data shows a prominent decay in the ratio over the
\XMM\ energy range.

%The orbital ratios, however show great difference. The 2000
%ratio plot has ratio values about 3 at softer energies, and
%gradually falls down to values about 2 with the increasing energy.
%The 2003 ratio plot begins with ratios as high as 5 and falls
%rapidly down to values about 1.5 as we get to the higher energies.
%%The effect of
%%absorbtion for the 2000 observation reaches well above 5 keV since
%%the ratios are as high as 2.
%%The high ratios out to 5 keV are more consistent with an
%%ionized absorber at work responsible for the spectral
%%difference over the orbital period.
So there is a clear difference between 2000 and 2003 states of the
source as far as absorption over the orbital cycle is concerned.
This most probably due to the changing geometry and ionization
structure of the absorbing material on the binary plane/accretion
disc.

The estimates from the model integrated fluxes yielded luminosities
of 6.7 $\times$ 10$^{31}$ and 5.8 $\times$ 10$^{31}$ erg s$^{-1}$
and accretion rates of 5.75 $\times$ 10$^{-11}$ and 5.0 $\times$
10$^{-11}$ M$_{\odot}$ yr$^{-1}$ for 2000 and 2003 data
respectively. The X-ray luminosities are lower but the accretion
rates are in accordance with the previous findings of Beuermann et
al. (2003) from optical data. We find that this region is located
around the accretion impact zone creating a sizeable absorbing bulge
in a similar fashion to Low-mass X-ray binary systems. This region
has a smaller neutral hydrogen column density by at least a factor
of four in the year 2000 and affects a smaller range of phases, but
it still exists.

\section{Summary and Conclusions}

The \XMM EPIC pn (0.3-10. keV) spectrum of the IP EX Hya together
with spin and orbital phase-resolved spectroscopy were thoroughly
investigated for the first time in this work. We have used two
different observations of EX Hya to look for spectral variations
over time. In both observations the mass accretion rate and
luminosities are compatible with those of the previous quiescent
observations, hence the source is in a quiescent state both in 2000
and 2003. The X-ray emission is modelled as arising from three
different components: two collisional equilibrium plasma emission
models around 0.6-0.8 keV and 1.3-1.7 keV and a cooling-flow plasma
emission model with a distribution of 3 to 33 keV for the 2000 state
and 8 to 61 keV for the 2003 state. We find very little change in
the spectral parameters of the two collisional equilibrium plasma
emission models over the orbital and spin cycles. In general, the
two observations of the source reveal spectral variations. For the
2000 data, we find neutral hydrogen column densities in a range
0.01-0.2$\times$ 10$^{22}$\ cm$^{-2}$ from spin phase-resolved
spectroscopy (with covering fractions ranging from 0.2-0.5), and
0.01-0.08$\times$ 10$^{22}$\ cm$^{-2}$ from orbital phase-resolved
spectroscopy (where covering fraction is 1.0). For the 2003 data,
the neutral hydrogen column densities are in a range
0.03-0.1$\times$ 10$^{22}$\ cm$^{-2}$ (with covering fractions of
0.6-1.0) in the spin resolved spectroscopy, and 0.02-0.15 $\times$
10$^{22}$\ cm$^{-2}$ in the orbital phase-resolved spectroscopy
(with covering fraction of 1.0). The highest $N_{\rm H}$ values
always occur either at the orbital or spin phase minima. There is a
definite difference in the absorber geometry for the spin
phase-resolved spectroscopy between the two epochs. In the year 2000
it shows a covering fraction of 0.2-0.5 whereas in the year 2003 it
shows a covering fraction of 0.6-1.0 (both years show that covering
fraction reduces towards spin maxima). This indicates a geometrical
change in the absorber from 2000 to 2003. The angle subtended by the
absorber is considerably less in 2000 and hence the accretion
curtain is reduced in size. Since we attained a model of only
neutral absorption, we can not comment much on the ionization state
changes of the absorber. We note that we do not rule out any
occultation effect or change in the occultation effects. The
absorption in the orbital plane changes by a factor of 2 from 2000
to 2003 and the absorption effects are more spread over the orbital
cycle in comparison with the 2000 data. Thus, the bulge at the
accretion impact zone is larger (and optically thicker) in 2003. We
also note that both the spin and orbital phase-resolved spectroscopy
of the 2003 data show a larger variation (3-25 keV) for the $LowT$
parameter of the cooling-flow plasma emission model as compared to
the 2000 state where this parameter stays around 6-12 keV over most
of the phase-resolved spectroscopy. Thus, the structure of the
cooling flow and the absorption effects are different in the two
different epochs of observations. This is not related to any mass
accretion rate difference, but only to a difference in the accretion
column temperature distribution structure and absorption.
%We detect
%The source clearly shows spectral modulation over the orbital phase.
%This modulation is clearly due to absorption since it can be seen
%that as the flux decreases with phase, the absorption increases.
%Spectral hardening at minimum phases also shows up indicating an
%existence of extra ionizing absorber.

We detect a 6.4 keV fluorescent Fe line at orbital phases 0.9-1.3
corresponding to the location of orbital minima in 2003 and it is
absent in the orbital phases 0.4 to 0.8. This feature can be
attributed to reflection from a region on the orbital  plane most
likely the bulge at the accretion impact zone and spread around the
disk covering a location including the eclipse phases. So the
absorber may be viewed in other wavelengths at around these phases.
We note that a 6.4 keV fluorescent Fe line is non-existent in the
orbital phase-resolved spectroscopy of the
 2000 data. This strongly indicates an accretion geometry
change and a change in the bulge size (impact zone) from year 2000
to 2003; the bulge is larger and more spread in 2003.

\section*{Acknowledgments}

The authors acknowledge support from T\"UB\.ITAK, The Scientific and
Technological Research Council of Turkey,  through project 108T735.

\appendix

\bsp

\newpage
\begin{table*}
\centering
\begin{minipage}{140mm}
\caption{Spectral parameters of the year 2000 and 2003 observations
in the 0.3-10 keV range. Both spectra were fitted with a composite
model of two collisional equilibrium plasma emission models at
different temperatures (MEKAL), two Gaussians centered at 6.7 keV,
and 6.9 keV, a variable cooling-flow plasma emission model
(VMCFLOW), a partial covering absorber model (PCFABS) and a simple
absorption model (TBABS). For the year 2000 spectrum, three more
Gaussians were added at 0.58 keV, 0.78 keV and 7.8 keV. For the year
2003 spectrum three more Gaussians were added at 6.4 keV, 0.58 keV
and 7.8 keV. $N_{\rm{H}}$ is the absorbing neutral hydrogen column
density, CoverFrac is the covering fraction of the absorber. The
given errors correspond to 2$\sigma$\ confidence level for a single
parameter. Gaussian line centers were fixed at their best fit values
and the iron abundance parameter in VMCFLOW was fixed to 0.2 to
improve the quality of the fits.}

\begin{tabular}{llrrr@{}}

\hline \hline

\multicolumn{1}{l}{Model}  & \multicolumn{1}{l}{Component}  &
\multicolumn{1}{r}{2000 Total Spectrum} &
\multicolumn{1}{r}{2003 Total Spectrum}\\

\hline

tbabs & $N_{\rm{H}}$ ($\times$ 10$^{22}$ atoms/cm$^{2}$) & 0.0097$^{+0.0007}_{-0.0007}$ & 0.0092$^{+0.0025}_{-0.0035}$ \\
pcfabs & $N_{\rm{H}}$ ($\times$ 10$^{22}$ atoms/cm$^{2}$) & 0.86$^{+0.17}_{-0.16}$ & 0.17$^{+0.06}_{-0.04}$ \\
& CoverFrac & 0.22$^{+0.01}_{-0.01}$ & 0.35$^{+0.7}_{-0.5}$ \\

MEKAL1 & kT & 0.66$^{+0.01}_{-0.01}$ & 0.63$^{+0.01}_{-0.01}$ \\
& norm ($\times$ 10$^{-3}$) & 4.8$^{+0.2}_{-0.2}$ & 2.9$^{+0.2}_{-0.1}$ \\

MEKAL2 & kT & 1.58$^{+0.05}_{-0.05}$ & 1.65$^{+0.07}_{-0.08}$ \\
& norm ($\times$ 10$^{-3}$) & 7.8$^{+0.7}_{-0.7}$ & 4.3$^{+0.5}_{-0.5}$\\

Gaussian1 (6.7 keV) & sigma (keV) & 0.070$^{+0.006}_{-0.005}$ & 0.076$^{+0.009}_{-0.010}$ \\
& norm ($\times$ 10$^{-4}$) & 2.8$^{+0.1}_{-0.1}$ & 2.6$^{+0.2}_{-0.2}$\\

Gaussian2 (6.9 keV) & sigma (keV) & 0.042$^{+0.016}_{-0.020}$ & 0.064$^{+0.035}_{-0.040}$ \\
& norm ($\times$ 10$^{-5}$) & 6.3$^{+0.6}_{-0.6}$ & 5.8$^{+1.1}_{-1.1}$ \\

Gaussian3 (6.4 keV) & sigma (keV) & N/A & 0.11$^{+0.04}_{-0.04}$ \\
& norm ($\times$ 10$^{-5}$) & N/A & 4.5$^{+1.1}_{-1.0}$ \\

Gaussian4 (7.8 keV) & sigma (keV) & 0.06$^{+0.06}_{-0.06}$ & 0.51$^{+0.29}_{-0.14}$ \\
& norm ($\times$ 10$^{-5}$) & 2.7$^{+1.0}_{-0.7}$ & 18.2$^{+7.6}_{-4.6}$ \\

Gaussian5 (0.58 keV) & sigma (keV) & 0 (frozen) & 0 (frozen)\\
& norm ($\times$ 10$^{-4}$) & 3.9$^{+0.6}_{-0.6}$ & 5.2$^{+0.8}_{-0.8}$ \\

Gaussian6 (0.78 keV) & sigma (keV) & 0 (frozen) & N/A \\
& norm ($\times$ 10$^{-4}$) & 5.0$^{+0.6}_{-0.6}$ & N/A \\

VMCFLOW & LowT & 3.4$^{+0.6}_{-0.6}$ & 9.0$^{+1.2}_{-1.0}$ \\
& HighT & 29.1$^{+3.6}_{-3.4}$ & 59.2$^{+2.4}_{-2.1}$ \\
& Norm ($\times$ 10$^{-8}$) & 5.9$^{+1.0}_{-0.7}$ & 2.9$^{+0.2}_{-0.1}$ \\

& $\chi^2_{\nu}$ & 1.70 (956 d.o.f.) & 1.66 (903 d.o.f.) \\

\hline
\end{tabular}
\end{minipage}
\end{table*}

\newpage

\begin{table*}
\centering
\begin{minipage}{140mm}
\caption{Parameters for the common Gaussian emission lines in 2000
and 2003 spectra. }

\begin{tabular}{llrrr@{}}

\hline \hline

\multicolumn{1}{l}{Emission Line}  & \multicolumn{1}{l}{Component} &
\multicolumn{1}{r}{2000 Spectrum} &
\multicolumn{1}{r}{2003 Spectrum}\\

\hline

Gaussian1 (6.7 keV) & peak energy (keV) & 6.7 (fixed) & 6.7 (fixed) \\
& sigma (keV) & 0.070$^{+0.006}_{-0.005}$ & 0.076$^{+0.009}_{-0.010}$ \\
& norm ($\times$ 10$^{-4}$) & 2.8$^{+0.1}_{-0.1}$ & 2.6$^{+0.2}_{-0.2}$\\

Gaussian2 (6.9 keV) & peak energy (keV) & 6.9 (fixed) & 6.9 (fixed) \\
& sigma (keV) & 0.042$^{+0.016}_{-0.020}$ & 0.064$^{+0.035}_{-0.040}$ \\
& norm ($\times$ 10$^{-5}$) & 6.3$^{+0.6}_{-0.6}$ & 5.8$^{+1.1}_{-1.1}$ \\

Gaussian3 (7.8 keV) & peak energy (keV) & 7.8$^{+0.3}_{-0.3}$ & 7.9$^{+0.8}_{-0.8}$ \\
& sigma (keV) & 0.059$^{+0.063}_{-0.059}$ & 0.51$^{+0.29}_{-0.14}$  \\
& norm ($\times$ 10$^{-5}$) & 2.6$^{+1.0}_{-0.7}$ &
18.2$^{+7.7}_{-4.6}$ \\

\hline

\end{tabular}
\end{minipage}
\end{table*}

\begin{table*}
\centering
\begin{minipage}{140mm}
%\begin{center}
%\label{1} 
\caption{Spectral parameters derived for the year 2000
observation at each orbital phase of 0.1 in the 0.3-10 keV range.
All the spectra were fitted with a composite model of two
collisional equilibrium plasma emission models at different
temperatures (MEKAL), 2 Gaussians centered at 6.7 keV and 6.9 keV, a
variable cooling-flow plasma emission model (VMCFLOW) and
photoelectric absorption of HI (WABS). The given errors correspond
to 2$\sigma$\ confidence level for a single parameter. Gaussian line
centers were fixed at their best fit values and the iron abundance
parameter in VMCFLOW was fixed to 0.2 to improve the quality of the
fits.}

\begin{tabular}{@{}llrrrrrrrrrrr@{}}
\hline \hline

\multicolumn{1}{l}{Model} & \multicolumn{1}{l}{Component}&
\multicolumn{1}{r}{0.1} & \multicolumn{1}{r}{0.2} &
\multicolumn{1}{r}{0.3} & \multicolumn{1}{r}{0.4} &
\multicolumn{1}{r}{0.5} \\

\hline

wabs & $N_{\rm{H}}$ ($\times$ $10^{-22}$) &
0.035$^{+0.002}_{-0.002}$ & 0.015$^{+0.002}_{-0.002}$ &
0.010$^{+0.002}_{-0.002}$ & 0.013$^{+0.002}_{-0.002}$ &
0.016$^{+0.002}_{-0.003}$ &
 \\ [1ex]

MEKAL1 & kT & 0.64$^{+0.01}_{-0.01}$ & 0.62$^{+0.02}_{-0.02}$ &
0.63$^{+0.01}_{-0.01}$ & 0.64$^{+0.01}_{-0.01}$ &
0.63$^{+0.01}_{-0.01}$ \\

& Norm ($\times$ $10^{-4}$) & 3.5$^{+0.1}_{-0.1}$ &
4.3$^{+0.2}_{-0.2}$ & 4.7$^{+0.2}_{-0.2}$ & 4.8$^{+0.2}_{-0.2}$ &
4.3$^{+0.2}_{-0.2}$ \\
[1ex]

MEKAL2 & kT & 1.57$^{+0.12}_{-0.14}$ & 1.37$^{+0.06}_{-0.07}$ &
1.56$^{+0.11}_{-0.13}$ & 1.59$^{+0.12}_{-0.17}$ &
1.60$^{+0.13}_{-0.17}$ \\

& Norm ($\times$ $10^{-4}$) & 5.9$^{+0.9}_{-0.9}$ &
5.6$^{+0.2}_{-0.2}$ & 7.9$^{+1.0}_{-1.1}$ & 7.5$^{+1.2}_{-1.3}$ &
6.4$^{+1.2}_{-1.2}$ \\
[1ex]

VMCFLOW & LowT  & 11.5$^{+0.7}_{-0.5}$ & 6.1$^{+3.0}_{-2.8}$ &
6.9$^{+0.4}_{-0.4}$ & 3.5$^{+0.6}_{-0.5}$ & 5.1$^{+0.9}_{-0.9}$ &
\\[2ex]

& HighT  & 14.7$^{+0.8}_{-1.2}$ & 18.2$^{+9.5}_{-3.1}$ &
18.6$^{+5.7}_{-3.4}$ & 32.8$^{+1.0}_{-0.9}$ & 29.5$^{+1.4}_{-1.2}$ &
\\

& Norm ($\times$ $10^{-9}$) & 39.4$^{+15.7}_{-11.8}$ &
12.2$^{+5.6}_{-5.6}$ & 14.4$^{+2.6}_{-3.1}$ & 6.1$^{+13.7}_{-1.6}$ &
6.2$^{+1.2}_{-0.7}$ &
\\[1ex]

Gaussian1 (6.7 keV) & $\sigma$ & 0.11$^{+0.02}_{-0.02}$ &
0.11$^{+0.03}_{-0.2}$ & 0.11$^{+0.04}_{-0.01}$ &
0.11$^{+0.03}_{-0.02}$ & 0.14$^{+0.03}_{-0.03}$ &
\\

& Norm ($\times$ $10^{-5}$) & 2.7$^{+0.5}_{-0.3}$ &
5.7$^{+0.4}_{-0.2}$ & 3.2$^{+0.5}_{-0.4}$ & 3.1$^{+0.4}_{-0.4}$ &
3.4$^{+0.6}_{-0.5}$ &
 \\[1ex]

Gaussian2 (6.9 keV) & $\sigma$ & 0 & 0 & 0 & 0 & 0 \\
& Norm ($\times$ $10^{-6}$) & 3.5$^{+1.9}_{-2.5}$ &
3.6$^{+2.0}_{-2.5}$ & 3.8$^{+2.0}_{-2.5}$ & 2.8$^{+2.0}_{-2.5}$ &
2.6$^{+3.0}_{-1.8}$ \\[1ex]

& $\chi^2_{\nu}$ (d.o.f) & 1.35 (510) & 1.31 (557 ) & 1.29 (574) &
1.29 (585) & 1.20 (548) &
\\[1ex]

\hline

\hline \hline

\multicolumn{1}{l}{Model} & \multicolumn{1}{l}{Component}&
\multicolumn{1}{r}{0.6} & \multicolumn{1}{r}{0.7} &
\multicolumn{1}{r}{0.8} & \multicolumn{1}{r}{0.9} &
\multicolumn{1}{r}{1.0} \\

\hline

wabs & $N_{\rm{H}}$ ($\times$ $10^{-22}$) &
0.010$^{+0.002}_{-0.002}$ & 0.016$^{+0.002}_{-0.002}$ &
0.018$^{+0.002}_{-0.002}$ & 0.027$^{+0.002}_{-0.002}$ &
0.077$^{+0.003}_{-0.003}$ \\ [1ex]

MEKAL1 & kT &  0.63$^{+0.02}_{-0.02}$ & 0.65$^{+0.01}_{-0.01}$ &
0.62$^{+0.02}_{-0.02}$ & 0.63$^{+0.01}_{-0.02}$ & 0.64$^{+0.02}_{-0.03}$ \\

& Norm ($\times$ $10^{-4}$) & 3.8$^{+0.1}_{-0.1}$ &
4.1$^{+0.2}_{-0.2}$ & 3.7$^{+0.2}_{-0.2}$ &
2.8$^{+0.1}_{-0.2}$ & 2.6$^{+0.2}_{-0.2}$\\
[1ex]

MEKAL2 & kT & 1.59$^{+0.13}_{-0.15}$ & 1.61$^{+0.12}_{-0.19}$ &
1.42$^{+0.16}_{-0.03}$ & 1.42$^{+0.09}_{-0.11}$ & 1.43$^{+0.14}_{-0.09}$ \\

& Norm ($\times$ $10^{-4}$) & 5.5$^{+1.0}_{-0.9}$ &
6.2$^{+1.3}_{-1.6}$ & 5.7$^{+1.0}_{-0.4}$ &
3.8$^{+0.8}_{-0.5}$ & 4.1$^{+0.9}_{-0.4}$ \\
[1ex]

VMCFLOW & LowT  & 7.8$^{+5.6}_{-2.5}$ & 7.4$^{+0.5}_{-0.5}$ &
9.2$^{+1.1}_{-0.7}$ &
9.2$^{+2.2}_{-1.4}$ & 9.9$^{+0.8}_{-0.9}$\\[2ex]

& HighT  & 22.6$^{+1.6}_{-1.0}$ & 15.4$^{+0.6}_{-0.7}$ &
21.7$^{+2.0}_{-1.0}$ &
43.9$^{+3.9}_{-2.6}$ & 22.7$^{+1.3}_{-1.2}$\\

& Norm ($\times$ $10^{-9}$) & 9.2$^{+9.0}_{-9.0}$ &
16.9$^{+6.6}_{-1.0}$ & 9.0$^{+5.0}_{-5.0}$ &
4.7$^{+0.5}_{-1.4}$ & 12.1$^{+2.5}_{-2.2}$ \\[1ex]

Gaussian1 (6.7 keV) & $\sigma$ & 0.09$^{+0.02}_{-0.02}$ &
0.06$^{+0.03}_{-0.04}$ & 0.10$^{+0.07}_{-0.02}$ &
0.12$^{+0.03}_{-0.02}$ & 0.09$^{+0.02}_{-0.02}$\\

& Norm ($\times$ $10^{-5}$) & 2.9$^{+0.4}_{-0.3}$ &
2.2$^{+0.4}_{-0.4}$ & 2.9$^{+0.2}_{-0.3}$
& 3.6$^{+0.4}_{-0.4}$ & 3.1$^{+0.4}_{-0.3}$ \\[1ex]

Gaussian2 (6.9 keV) & $\sigma$ & 0 & 0 & 0 & 0 & 0 \\

& Norm ($\times$ $10^{-6}$) & 5.5$^{+2.1}_{-2.2}$ &
5.1$^{+1.2}_{-2.1}$ &
2.2$^{+1.9}_{-2.2}$ & 4.2$^{+2.1}_{-1.2}$ & 3.1$^{+2.3}_{-1.8}$\\[1ex]

& $\chi^2_{\nu}$ (d.o.f) & 1.23 (522) & 1.23 (538) & 1.33 (531) & 1.40 (501) & 1.30 (531)\\[1ex]

\hline
\end{tabular}
%\end{center}
\end{minipage}
\end{table*}

\begin{table*}
\centering
\begin{minipage}{140mm}
%\label{2} 
\caption{ Spectral parameters derived for the year 2003
observation at each orbital phase of 0.1 in the 0.3-10 keV range.
All the spectra were fitted with a composite model of two
collisional equilibrium plasma emission models at different
temperatures (MEKAL), 2 Gaussians centered at 6.7 keV and 6.9 keV, a
variable cooling-flow plasma emission model (VMCFLOW) and
photoelectric absorption of HI (WABS). In some phases another
Gaussian at 6.4 keV was added to the model. The given errors
correspond to 2$\sigma$\ confidence level for a single parameter.
Gaussian line centers were fixed at their best fit values and the
iron abundance parameter in VMCFLOW was fixed to 0.2 to improve the
quality of the fits. }

%\begin{center}
\begin{tabular}{@{}llrrrrrrrrrrr@{}}
\hline \hline

\multicolumn{1}{l}{Model} & \multicolumn{1}{l}{Component}&
\multicolumn{1}{r}{0.1} & \multicolumn{1}{r}{0.2} &
\multicolumn{1}{r}{0.3} & \multicolumn{1}{r}{0.4} &
\multicolumn{1}{r}{0.5}\\

\hline

wabs & $N_{\rm{H}}$ ($\times$ $10^{-22}$) &
0.142$^{+0.006}_{-0.006}$ & 0.046$^{+0.003}_{-0.003}$ &
0.026$^{+0.003}_{-0.003}$ &
0.020$^{+0.002}_{-0.002}$ & 0.024$^{+0.002}_{-0.002}$  \\[1ex]

MEKAL1 & kT & 0.63$^{+0.04}_{-0.06}$ & 0.65$^{+0.02}_{-0.02}$ &
0.61$^{+0.03}_{-0.03}$ & 0.65$^{+0.02}_{-0.02}$ &
0.64$^{+0.02}_{-0.02}$\\

& Norm ($\times$ $10^{-4}$) & 1.2$^{+0.1}_{-0.2}$ &
2.6$^{+0.2}_{-0.2}$ & 2.6$^{+0.3}_{-0.2}$ & 3.3$^{+0.2}_{-0.2}$ &
2.7$^{+0.2}_{-0.1}$ \\
[1ex]

MEKAL2 & kT & 1.53$^{+0.18}_{-0.19}$ & 1.80$^{+0.26}_{-0.13}$ &
1.48$^{+0.12}_{-0.08}$ & 1.72$^{+0.09}_{-0.09}$ &
1.65$^{+0.14}_{-0.12}$\\

& Norm ($\times$ $10^{-4}$) & 2.3$^{+0.6}_{-0.6}$ &
4.3$^{+1.3}_{-0.8}$ & 3.9$^{+0.9}_{-0.8}$ & 5.2$^{+0.9}_{-0.8}$ &
3.6$^{+0.8}_{-0.8}$ \\
[1ex]

VMCFLOW & LowT  & 25.7$^{+1.0}_{-5.0}$ & 10.8$^{+19.5}_{-6.6}$ &
4.3$^{+1.7}_{-0.9}$ & 5.7$^{+3.4}_{-1.6}$ & 4.1$^{+1.4}_{-1.9}$ \\

& HighT  & 79.9$^{<}_{-19.4}$ & 60.8$^{+12.3}_{-20.2}$ &
57.2$^{+4.4}_{-2.9}$ & 58.2$^{+3.2}_{-3.1}$ & 60.4$^{+3.8}_{-8.5}$ \\

& Norm ($\times$ $10^{-9}$) & 4.1$^{+1.4}_{-0.1}$ &
4.1$^{+1.1}_{-0.8}$ & 3.8$^{+0.6}_{-0.4}$ & 2.7$^{+1.0}_{-0.1}$ &
3.4$^{+0.3}_{-0.2}$\\
[1ex]

Gaussian1 (6.7 keV) & $\sigma$ & 0.07$^{+0.03}_{-0.03}$ &
0.10$^{+0.06}_{-0.04}$ & 0.10$^{+0.09}_{-0.04}$ &
0.25$^{+0.08}_{-0.05}$ & 0.19$^{+0.06}_{-0.05}$ \\

& Norm ($\times$ $10^{-5}$) & 3.1$^{+0.6}_{-0.5}$ &
2.3$^{+1.2}_{-0.5}$ & 1.4$^{+0.9}_{-0.6}$ & 3.8$^{+0.9}_{-0.8}$ &
3.2$^{+0.7}_{-0.7}$ \\
[1ex]

Gaussian2 (6.9 keV)& $\sigma$ &  0 & 0 & 0 & 0 & 0 \\

& Norm ($\times$ $10^{-6}$) & 2.1$^{+3.5}_{-2.1}$ &
5.1$^{+3.7.0}_{-5.1}$ & 1.5$^{+0.5}_{-0.3}$ &  $<$ 0.000037 &  $<$ 0.000021 \\
[1ex]

Gaussian3 (6.4 keV) & $\sigma$ & 0.08$^{+0.11}_{-0.08}$ &
0.09$^{+0.44}_{-0.09}$ & 0.16$^{+0.08}_{-0.06}$ & 0 & 0 \\

& Norm ($\times$ $10^{-6}$) & 8.7$^{+4.5}_{-5.2}$ &
6.3$^{+4.7}_{-3.2}$ & 14.9$^{+4.8}_{-7.3}$ &  $<$ 0.0018 & $<$ 0.000021 \\
[1ex]

& $\chi^2_{\nu}$ (d.o.f.) & 1.27 (254) & 1.22 (363) & 1.40
(380) & 1.26 (386) & 1.43 (374) \\
[1ex]

\hline

\hline \hline

\multicolumn{1}{l}{Model} & \multicolumn{1}{l}{Component}&
\multicolumn{1}{r}{0.6} & \multicolumn{1}{r}{0.7} &
\multicolumn{1}{r}{0.8} & \multicolumn{1}{r}{0.9} &
\multicolumn{1}{r}{1.0} \\

\hline

wabs & $N_{\rm{H}}$ ($\times$ $10^{-22}$) &
0.025$^{+0.003}_{-0.002}$ & 0.022$^{+0.002}_{-0.002}$ &
0.020$^{+0.002}_{-0.002}$ &
0.022$^{+0.002}_{-0.002}$ & 0.085$^{+0.004}_{-0.004}$ \\[1ex]

MEKAL1 & kT & 0.65$^{+0.2}_{-0.2}$ & 0.63$^{+0.02}_{-0.03}$ &
0.63$^{+0.02}_{-0.02}$ &
0.62$^{+0.03}_{-0.03}$ & 0.67$^{+0.07}_{-0.08}$\\

& Norm ($\times$ $10^{-4}$) & 2.7$^{+0.2}_{-0.2}$ &
3.0$^{+0.2}_{-0.2}$ &
3.1$^{+0.2}_{-0.2}$ & 2.5$^{+0.2}_{-0.3}$ & 1.0$^{+0.2}_{-0.2}$ \\
[1ex]

MEKAL2 & kT & 1.64$^{+0.11}_{-0.13}$ & 1.36$^{+0.07}_{-0.07}$ &
1.66$^{+0.12}_{-0.14}$ &
1.35$^{+0.08}_{-0.07}$ & 1.27$^{+0.16}_{-0.14}$\\

& Norm ($\times$ $10^{-4}$) & 3.9$^{+0.8}_{-0.8}$ &
2.9$^{+0.6}_{-0.4}$ &
4.3$^{+0.9}_{-0.9}$ & 2.4$^{+0.7}_{-0.5}$ & 1.2$^{+0.3}_{-0.4}$ \\
[1ex]

VMCFLOW & LowT  & 7.3$^{+3.4}_{-2.6}$ & 3.5$^{+1.3}_{-1.0}$ &
3.1$^{+1.3}_{-1.4}$ &
6.9$^{+0.8}_{-2.0}$ & 24.0$^{+0.8}_{-1.0}$  \\

& HighT  & 73.5$^{+4.5}_{-12.5}$ & 62.0$^{+3.6}_{-3.3}$ &
62.0$^{+3.7}_{-3.1}$ &
79.9$^{<}_{-11.8}$ & 79.9$^{<}_{-14.9}$ \\

& Norm ($\times$ $10^{-9}$) & 3.4$^{+0.5}_{-0.4}$  &
3.2$^{+0.5}_{-0.4}$ & 3.4$^{+0.9}_{-0.7}$ &
3.1$^{+0.3}_{-0.2}$ & 4.2$^{+0.8}_{-0.1}$ \\
[1ex]

Gaussian1 (6.7 keV) & $\sigma$ & 0.15$^{+0.06}_{-0.03}$ &
0.20$^{+0.05}_{-0.04}$ &
0.22$^{+0.05}_{-0.04}$ & 0.06$^{+0.04}_{-0.03}$ & 0.08$^{+0.04}_{-0.04}$ \\

& Norm ($\times$ $10^{-5}$) & 3.1$^{+0.5}_{-0.8}$ &
4.4$^{+0.8}_{-0.7}$ & 4.1$^{+0.4}_{-0.7}$ &
2.4$^{+0.5}_{-0.8}$ & 2.2$^{+0.6}_{-0.5}$ \\
[1ex]

Gaussian2 (6.9 keV)& $\sigma$ &  0 & 0 & 0 & 0 & 0\\
& Norm ($\times$ $10^{-6}$) & $<$ 0.00016 & $<$ 2.3 & $<$ 1.3 &
8.3$^{+3.8}_{-3.8}$ & 7.9$^{+3.7}_{-4.1}$ \\
[1ex]

Gaussian3 (6.4 keV) & $\sigma$ & 0 & 0 & 0 &
0.0$^{+0.3.2}_{>}$ & 0.10$^{+0.06}_{-0.05}$\\

& Norm ($\times$ $10^{-6}$) & $<$ 0.84 & $<$ 0.0022 & $<$ 0.0013 &
5.2$^{+8.8}_{-3.0}$ & 12.1$^{+4.9}_{-5.0}$ \\ [1ex]

& $\chi^2_{\nu}$ (d.o.f.) & 1.17 (385) & 1.41 (392) & 1.34 (393) & 1.44 (376) & 1.65 (273)\\
[1ex]

\hline
\end{tabular}
%\end{center}
\end{minipage}
\end{table*}

\begin{table*}
\centering
\begin{minipage}{140mm}
%\label{3} 
\caption{ Spectral parameters derived for the year 2003
observation at each spin phase of 0.1 in the 0.3-10 keV range. All
the spectra were fitted with a composite model of two collisional
equilibrium plasma emission models at different temperatures
(MEKAL), a Gaussian centered at 6.7 keV, a variable cooling-flow
plasma emission model (VMCFLOW) and a partial covering absorber
model (PCFABS). The given errors correspond to 2$\sigma$\ confidence
level for a single parameter. Gaussian line centers were fixed at
their best fit values and the iron abundance parameter in VMCFLOW
was fixed to 0.2 to improve the quality of the fits. }

%\begin{center}
\begin{tabular}{@{}llrrrrrrrrrrr@{}}
\hline \hline

\multicolumn{1}{l}{Model} & \multicolumn{1}{l}{Component}&
\multicolumn{1}{r}{0.1} & \multicolumn{1}{r}{0.2} &
\multicolumn{1}{r}{0.3} & \multicolumn{1}{r}{0.4} &
\multicolumn{1}{r}{0.5} \\

\hline

pcfabs & $N_{\rm{H}}$ ($\times$ $10^{-22}$) &
0.027$^{+0.003}_{-0.029}$ & 0.044$^{+0.005}_{-0.006}$ &
0.093$^{+0.044}_{-0.039}$ &
0.051$^{+0.004}_{-0.006}$ & 0.109$^{+0.013}_{-0.014}$ \\

& CoverFrac & 1.00$^{<}_{-0.16}$ & 0.71$^{+0.08}_{-0.07}$ &
0.68$^{+0.08}_{-0.08}$ & 0.92$^{+0.06}_{-0.06}$ &
0.55$^{+0.09}_{-0.07}$ \\
[1ex]

MEKAL1 & kT & 0.52$^{+0.1}_{-0.05}$ & 0.63$^{+0.02}_{-0.03}$ &
0.60$^{+0.03}_{-0.04}$ & 0.62$^{+0.03}_{-0.04}$ &
0.62$^{+0.02}_{-0.04}$ \\

& Norm ($\times$ $10^{-4}$) & 1.8$^{+0.5}_{-0.2}$ &
2.2$^{+0.2}_{-0.2}$ & 1.8$^{+0.2}_{-0.2}$ & 1.9$^{+0.2}_{-0.1}$ &
3.2$^{+0.8}_{-0.5}$ \\
[1ex]

MEKAL2 & kT & 1.10$^{+0.28}_{-0.18}$ & 1.4$^{+0.2}_{-0.1}$ &
1.29$^{+0.10}_{-0.09}$ & 1.58$^{+0.08}_{-0.23}$ &
1.46$^{+0.18}_{-0.10}$ \\

& Norm ($\times$ $10^{-4}$) & 1.9$^{+0.2}_{-0.5}$ &
1.7$^{+0.7}_{-0.4}$ & 2.2$^{+0.4}_{-0.4}$ & 3.3$^{+0.7}_{-0.8}$ &
2.8$^{+0.8}_{-0.6}$ \\
[1ex]

VMCFLOW & LowT & 24.1$^{+0.4}_{-3.7}$ & 26.4$^{+0.9}_{-0.9}$ &
22.7$^{+0.4}_{-2.7}$ & 18.1$^{+13.1}_{-4.2}$ &
6.9$^{+2.1}_{-1.4}$ \\

& HighT  & 79.9$^{<}_{-12.3}$ & 79.9$^{<}_{-3.8}$ &
79.9$^{<}_{-11.7}$ & 79.9$^{<}_{-18.1}$ & 79.9$^{<}_{-10.3}$ \\

& Norm ($\times$ $10^{-9}$) & 3.8$^{+0.7}_{-0.1}$ &
3.8$^{+0.7}_{-0.1}$ & 3.7$^{+0.6}_{-0.1}$ & 3.7$^{+1.8}_{-0.1}$ &
3.2$^{+1.0}_{-1.0}$ \\
[1ex]

Gaussian1 (6.7 keV) & $\sigma$ & 0.18$^{+0.7}_{-0.5}$ &
0.19$^{+0.06}_{-0.04}$ & 0.18$^{+0.04}_{-0.03}$ &
0.20$^{+0.04}_{-0.04}$ &
0.14$^{+0.07}_{-0.04}$ \\

& Norm ($\times$ $10^{-5}$) & 3.6$^{+0.8}_{-0.6}$ &
3.8$^{+0.8}_{-0.6}$ & 4.2$^{+0.7}_{-0.6}$ & 4.0$^{+0.6}_{-0.7}$ &
2.4$^{+0.3}_{-0.4}$ \\
[1ex]

& $\chi^2_{\nu}$ (d.o.f.)& 1.27 (333) & 1.21 (312) & 1.17 (329) &
1.24 (338) & 1.35 (360) \\
[1ex]

\hline

\hline \hline

\multicolumn{1}{l}{Model} & \multicolumn{1}{l}{Component}&
\multicolumn{1}{r}{0.6} & \multicolumn{1}{r}{0.7} &
\multicolumn{1}{r}{0.8} & \multicolumn{1}{r}{0.9} &
\multicolumn{1}{r}{1.0} \\

\hline

pcfabs & $N_{\rm{H}}$ ($\times$ $10^{-22}$) &
0.055$^{+0.004}_{-0.005}$ & 0.073$^{+0.007}_{-0.017}$ &
0.069$^{+0.043}_{-0.034}$ & 0.028$^{+0.014}_{-0.002}$ & 0.061$^{+0.006}_{-0.010}$ \\

& CoverFrac & 0.77$^{+0.08}_{-0.13}$ & 0.64$^{+0.07}_{-0.10}$ &
0.48$^{+0.29}_{-0.12}$ &
1.0$^{<}_{-0.14}$ & 0.60$^{+0.13}_{-0.13}$ \\
[1ex]

MEKAL1 & kT & 0.64$^{+0.02}_{-0.02}$ & 0.64$^{+0.02}_{-0.02}$ &
0.63$^{+0.01}_{-0.01}$ & 0.64$^{+0.02}_{-0.02}$ & 0.64$^{+0.02}_{-0.03}$\\

& Norm ($\times$ $10^{-4}$) & 2.6$^{+0.2}_{-0.2}$ &
3.4$^{+0.3}_{-0.2}$ &
3.6$^{+0.2}_{-0.2}$ & 3.1$^{+0.2}_{-0.2}$ & 2.6$^{+0.2}_{-0.1}$ \\
[1ex]

MEKAL2 & kT & 1.67$^{+0.12}_{-0.10}$ & 1.72$^{+0.10}_{-0.13}$ &
1.92$^{+0.18}_{-0.15}$ & 1.66$^{+0.18}_{-0.22}$ & 1.63$^{+0.15}_{-0.16}$\\

& Norm ($\times$ $10^{-4}$) & 3.8$^{+0.9}_{-0.8}$ &
4.7$^{+1.3}_{-1.0}$ &
7.4$^{+0.8}_{-0.7}$ & 4.8$^{+1.2}_{-1.0}$ & 3.1$^{+0.9}_{-0.8}$ \\
[1ex]

VMCFLOW & LowT & 4.1$^{+1.0}_{-1.0}$ & 3.9$^{+0.6}_{-0.5}$ &
3.4$^{+1.6}_{-1.0}$ & 6.2$^{+3.0}_{-1.9}$ & 8.2$^{+3.3}_{-3.7}$ \\

& HighT  & 56.7$^{+9.0}_{-8.0}$ & 43.7$^{+18.5}_{-15.7}$ &
59.0$^{+1.8}_{-2.9}$ & 46.4$^{+12.7}_{-11.2}$ & 61.3$^{+3.2}_{-10.3}$ \\

& Norm ($\times$ $10^{-9}$) & 3.6$^{+0.7}_{-0.6}$ &
4.2$^{+0.6}_{-0.5}$ &
3.8$^{+0.7}_{-0.6}$ & 4.7$^{+0.9}_{-0.7}$ & 4.1$^{+1.0}_{-0.9}$ \\
[1ex]

Gaussian1 (6.7 keV) & $\sigma$ & 0.17$^{+0.07}_{-0.04}$ &
0.31$^{+0.08}_{-0.06}$ &
0.15$^{+0.09}_{-0.04}$ & 0.19$^{+0.04}_{-0.04}$ & 0.17$^{+0.05}_{-0.04}$ \\

& Norm ($\times$ $10^{-5}$) & 3.6$^{+0.08}_{-0.06}$ &
3.9$^{+0.8}_{-0.7}$ &
3.5$^{+1.0}_{-0.6}$ & 4.1$^{+0.7}_{-0.7}$ & 3.8$^{+0.7}_{-0.6}$ \\
[1ex]

& $\chi^2_{\nu}$ (d.o.f.)& 1.27 (374) & 1.46 (396) & 1.39 (416) & 1.39 (416) & 1.55 (382)\\
[1ex]

\hline
\end{tabular}
%\end{center}
\end{minipage}
\end{table*}

\begin{table*}
\centering
\begin{minipage}{140mm}
%\label{4} 
\caption{ Spectral parameters derived for the year 2000
observation at each spin phase of 0.1 in the 0.3-10 keV range. All
the spectra were fitted with a composite model of two collisional
equilibrium plasma emission models at different temperatures
(MEKAL), 2 Gaussians centered at 6.7 keV and 6.9 keV, a variable
cooling-flow plasma emission model (VMCFLOW) and a partial covering
absorber model (PCFABS). The given errors correspond to 2$\sigma$\
confidence level for a single parameter. Gaussian line centers were
fixed at their best fit values and the iron abundance parameter in
VMCFLOW was fixed to 0.2 to improve the quality of the fits. }

%\begin{center}
\begin{tabular}{@{}llrrrrrrrrrrr@{}}
\hline \hline

\multicolumn{1}{l}{Model} & \multicolumn{1}{l}{Component}&
\multicolumn{1}{r}{0.1} & \multicolumn{1}{r}{0.2} &
\multicolumn{1}{r}{0.3} & \multicolumn{1}{r}{0.4} &
\multicolumn{1}{r}{0.5} \\

\hline

pcfabs & $N_{\rm{H}}$ ($\times$ $10^{-22}$) & 0.17$^{+0.16}_{-0.09}$
& 0.06$^{+0.12}_{-0.04}$ & 0.52$^{+0.20}_{-0.25}$ &
0.44$^{+0.20}_{-0.17}$ & 0.20$^{+0.08}_{-0.07}$ \\

& CoverFrac & 0.24$^{+0.06}_{-0.03}$ & 0.31$^{+0.68}_{-0.17}$ &
0.42$^{+0.04}_{-0.04}$ & 0.49$^{+0.04}_{-0.02}$ &
0.43$^{+0.07}_{-0.03}$
\\ [1ex]

MEKAL1 & kT & 0.62$^{+0.2}_{-0.2}$ & 0.62$^{+0.02}_{-0.02}$ &
0.61$^{+0.03}_{-0.02}$ & 0.62$^{+0.01}_{-0.01}$ &
0.63$^{+0.01}_{-0.03}$ \\

& Norm ($\times$ $10^{-4}$) & 4.6$^{+0.4}_{-0.3}$ &
3.5$^{+0.3}_{-0.2}$ & 4.1$^{+0.7}_{-0.4}$ & 4.4$^{+0.7}_{-0.7}$ &
3.2$^{+0.3}_{-0.4}$ \\
[1ex]

MEKAL2 & kT & 1.41$^{+0.21}_{-0.07}$ & 1.42$^{+0.20}_{-0.07}$ &
1.41$^{+0.22}_{-0.08}$ & 1.58$^{+0.13}_{-0.16}$ &
1.47$^{+0.17}_{-0.13}$ \\

& Norm ($\times$ $10^{-4}$) & 5.4$^{+1.2}_{-0.6}$ &
4.6$^{+1.1}_{-0.5}$ & 5.7$^{+1.1}_{-1.2}$ & 6.9$^{+1.8}_{-2.4}$ &
4.7$^{+0.9}_{-0.8}$ \\
[1ex]

VMCFLOW & LowT  & 4.2$^{+1.3}_{-0.6}$ & 11.7$^{+3.5}_{-1.9}$ &
9.6$^{+2.2}_{-1.5}$ & 7.1$^{+1.2}_{-1.2}$ & 5.5$^{+1.0}_{-0.6}$ \\

& HighT  & 33.8$^{+1.3}_{-1.2}$ & 45.7$^{+5.2}_{-3.3}$ &
45.7$^{+3.8}_{-3.2}$ & 39.4$^{+1.2}_{-2.6}$ &
35.7$^{+2.0}_{-1.3}$ \\

& Norm ($\times$ $10^{-9}$) &  4.6$^{+5.5}_{-1.1}$ &
4.8$^{+1.1}_{-0.8}$ & 4.6$^{+1.1}_{-0.8}$ & 4.9$^{+1.6}_{-0.8}$ & 5.2$^{+2.2}_{-2.2}$ \\
[1ex]

Gaussian1 (6.7 keV) & $\sigma$ & 0.11$^{+0.03}_{-0.01}$ &
0.12$^{+0.05}_{-0.01}$ & 0.10$^{+0.04}_{-0.02}$ &
0.12$^{+0.04}_{-0.02}$ & 0.12$^{+0.03}_{-0.03}$ & \\

& Norm ($\times$ $10^{-5}$) & 2.8$^{+0.4}_{-0.4}$ &
3.2$^{+0.5}_{-0.3}$ & 3.2$^{+0.4}_{-0.2}$ & 3.0$^{+0.5}_{-0.4}$ & 3.1$^{+0.4}_{-0.2}$ \\
[1ex]

Gaussian2 (6.9 keV) & $\sigma$ & 0 & 0 & 0 & 0 & 0 \\

& Norm ($\times$ $10^{-6}$) & 5.5$^{+2.1}_{-2.3}$ &
3.4$^{+2.1}_{-2.8}$ & 4.0$^{+2.0}_{-2.4}$ & 0.9$^{+2.2}_{-0.7}$ &
0.9$^{+2.0}_{-0.7}$ \\
[1ex]

& $\chi^2_{\nu}$ & 1.27 (539) & 1.26 (493) & 1.28 (494) & 1.28 (501) & 1.23 (494) \\
[1ex]

\hline

\hline \hline

\multicolumn{1}{l}{Model} & \multicolumn{1}{l}{Component}&
\multicolumn{1}{r}{0.6} & \multicolumn{1}{r}{0.7} &
\multicolumn{1}{r}{0.8} & \multicolumn{1}{r}{0.9} &
\multicolumn{1}{r}{1.0} \\

\hline

pcfabs & $N_{\rm{H}}$ ($\times$ $10^{-22}$) & 0.12$^{+0.06}_{-0.04}$
& 0.20$^{+0.22}_{-0.11}$ &
0.12$^{+0.18}_{-0.08}$ & 0.12$^{+0.07}_{-0.07}$ & 0.10$^{+0.10}_{-0.06}$\\

& CoverFrac & 0.46$^{+0.10}_{-0.6}$ & 0.24$^{+0.07}_{-0.03}$ &
0.16$^{+0.11}_{-0.04}$ & 0.22$^{+0.13}_{-0.04}$ &
0.29$^{+0.28}_{-0.07}$
\\ [1ex]

MEKAL1 & kT & 0.61$^{+0.02}_{-0.02}$ & 0.61$^{+0.02}_{-0.02}$ &
0.62$^{+0.02}_{-0.02}$ &
0.64$^{+0.01}_{-0.01}$ & 0.64$^{+0.01}_{-0.01}$ \\

& Norm ($\times$ $10^{-4}$) & 3.7$^{+0.3}_{-0.2}$ &
4.5$^{+0.4}_{-0.3}$ & 4.9$^{+0.3}_{-0.3}$ &
6.0$^{+0.3}_{-0.3}$ & 6.0$^{+0.4}_{-0.2}$ \\
[1ex]

MEKAL2 & kT & 1.36$^{+0.07}_{-0.08}$ & 1.40$^{+0.10}_{-0.06}$ &
1.42$^{+0.15}_{-0.06}$ &
1.57$^{+0.12}_{-0.16}$ & 1.71$^{+0.10}_{-0.12}$ \\

& Norm ($\times$ $10^{-4}$) & 4.5$^{+0.6}_{-0.6}$ &
6.1$^{+0.4}_{-0.6}$ & 7.0$^{+0.6}_{-0.6}$ &
8.1$^{+1.5}_{-1.5}$ & 9.3$^{+1.7}_{-1.6}$ \\
[1ex]

VMCFLOW & LowT  & 2.6$^{+0.6}_{-0.5}$ & 3.4$^{+0.5}_{-0.4}$ &
3.7$^{+0.6}_{-0.4}$ & 2.5$^{+0.4}_{-0.4}$ & 2.7$^{+0.4}_{-0.4}$ \\

& HighT  & 34.8$^{+1.1}_{-0.9}$ & 21.5$^{+0.7}_{-0.6}$ &
21.6$^{+0.7}_{-0.6}$ &
22.7$^{+0.5}_{-0.5}$ & 22.8$^{+0.5}_{-0.5}$ \\

& Norm ($\times$ $10^{-9}$) & 4.4$^{+6.0}_{-0.5}$ &
7.7$^{+2.5}_{-2.5}$ &
8.9$^{+5.2}_{-2.8}$ & 8.8$^{+4.5}_{-1.4}$ & 7.9$^{+5.7}_{-1.6}$ \\
[1ex]

Gaussian1 (6.7 keV) & $\sigma$ & 0.10$^{+0.04}_{-0.02}$ &
0.10$^{+0.03}_{-0.02}$ &
0.11$^{+0.01}_{-0.02}$ & 0.10$^{+0.02}_{-0.02}$ & 0.11$^{+0.03}_{-0.01}$ \\

& Norm ($\times$ $10^{-5}$) & 3.1$^{+0.5}_{-0.3}$ &
2.7$^{+0.4}_{-0.3}$ &
3.2$^{+0.4}_{-0.4}$ & 3.2$^{+0.4}_{-0.4}$ & 3.0$^{+0.4}_{-0.4}$ \\
[1ex]

Gaussian2 (6.9 keV) & $\sigma$ & 0 & 0 & 0 & 0 & 0 \\

& Norm ($\times$ $10^{-6}$) & 4.2$^{+1.9}_{-2.5}$ &
4.6$^{+2.0}_{-2.2}$ & 5.6$^{+2.2}_{-2.4}$ &
5.3$^{+2.1}_{-2.3}$ & 5.0$^{+2.4}_{-2.6}$ \\
[1ex]

& $\chi^2_{\nu}$ & 1.31 (521) & 1.39 (527) & 1.22 (581) & 1.10 (600) & 1.24 (583) \\
[1ex]

\hline
\end{tabular}
%\end{center}
\end{minipage}
\end{table*}
\vspace{0.3cm}

\begin{figure*}
\centerline{
\includegraphics[scale=0.5,angle=270]{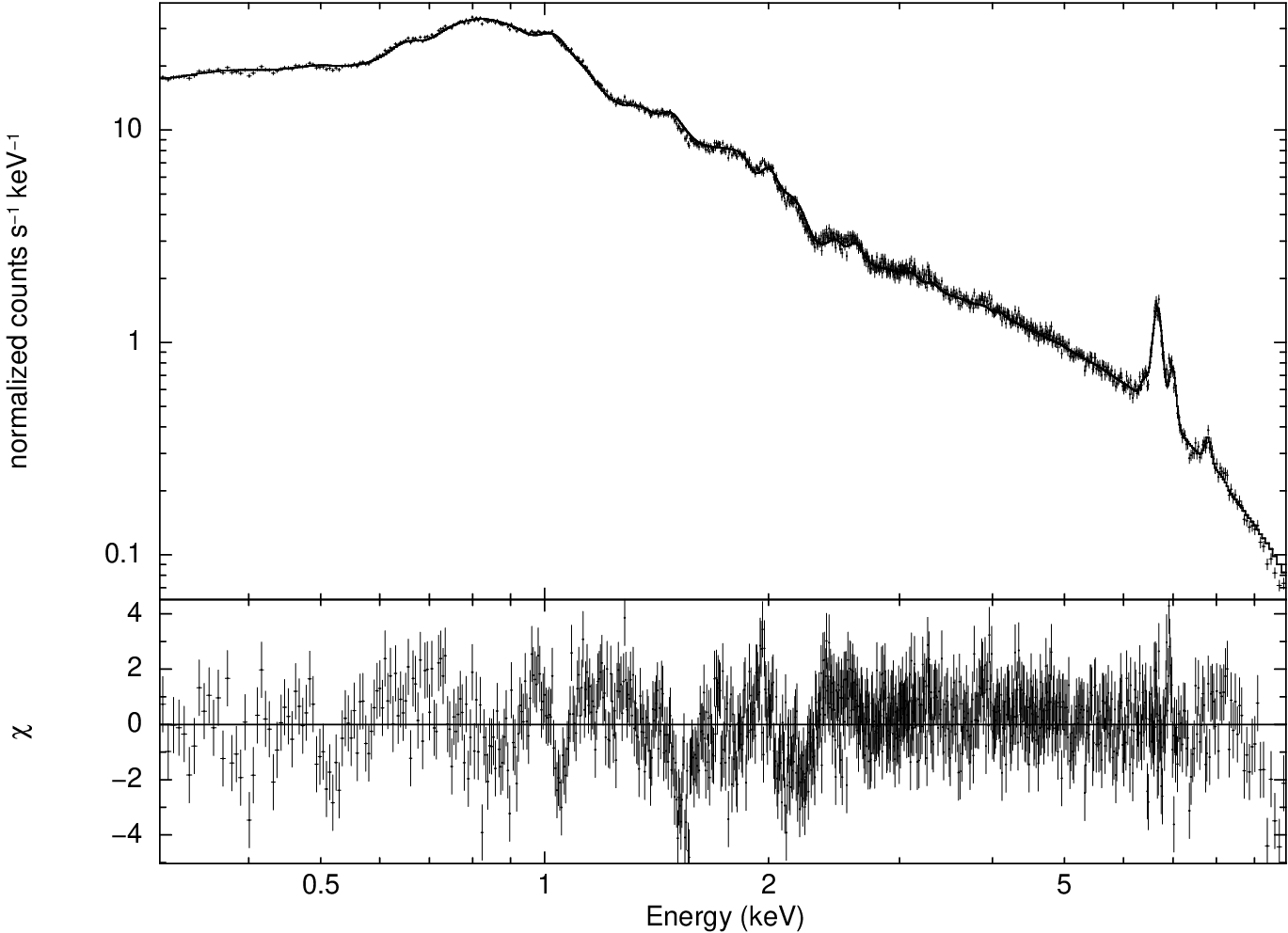}}
\centerline{
\includegraphics[scale=0.5,angle=270]{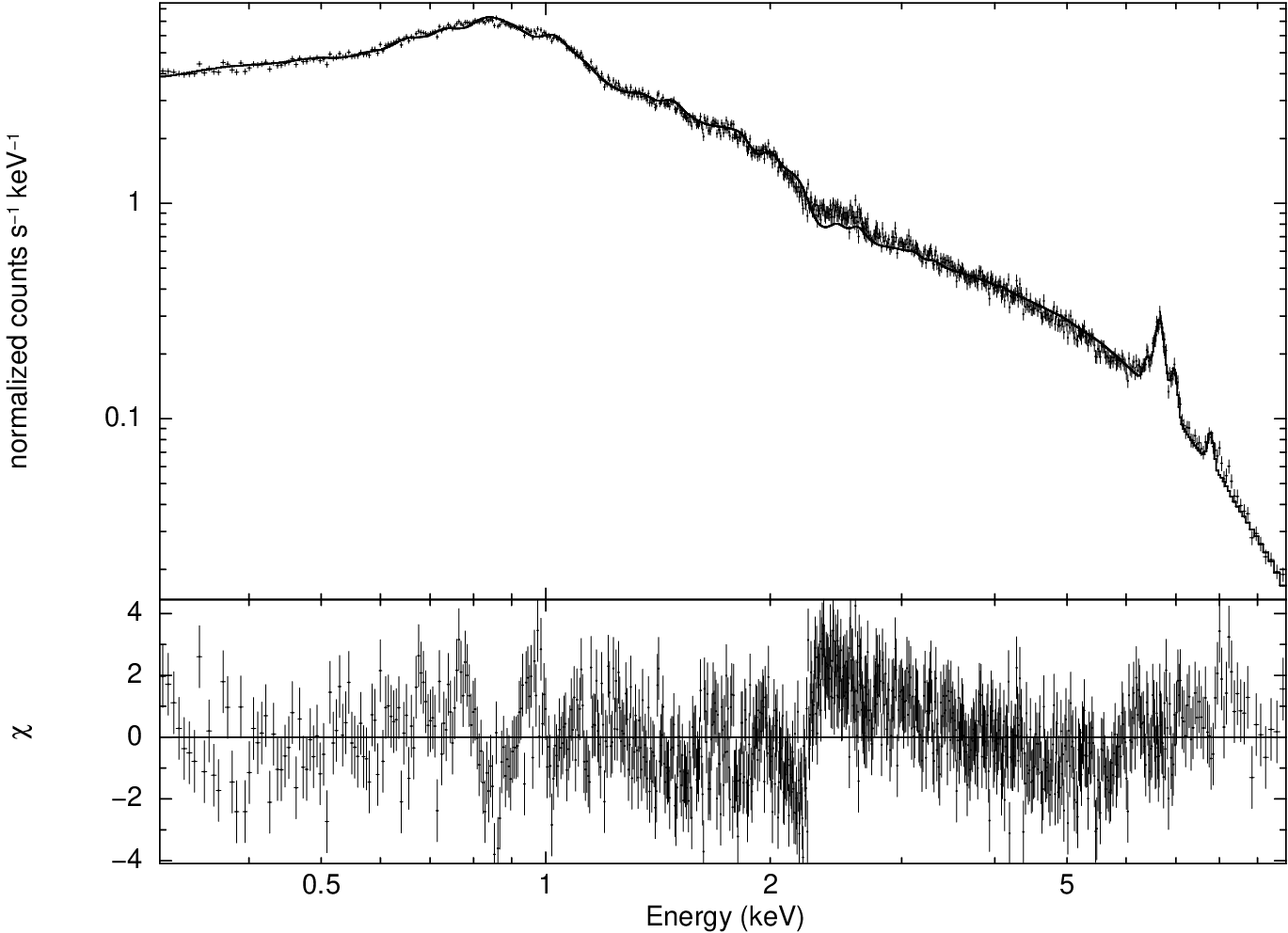}}
\caption{The top panel shows the total EPIC pn spectrum of the year
2000 data (see $\textbf{Table 1}$ for details). The bottom panel
shows the total EPIC pn spectrum of the year 2003 data (see
$\textbf{Table 1}$ for details).}
\end{figure*}

\newpage

\begin{figure*}
\centerline{
\includegraphics[scale=0.35,angle=270]{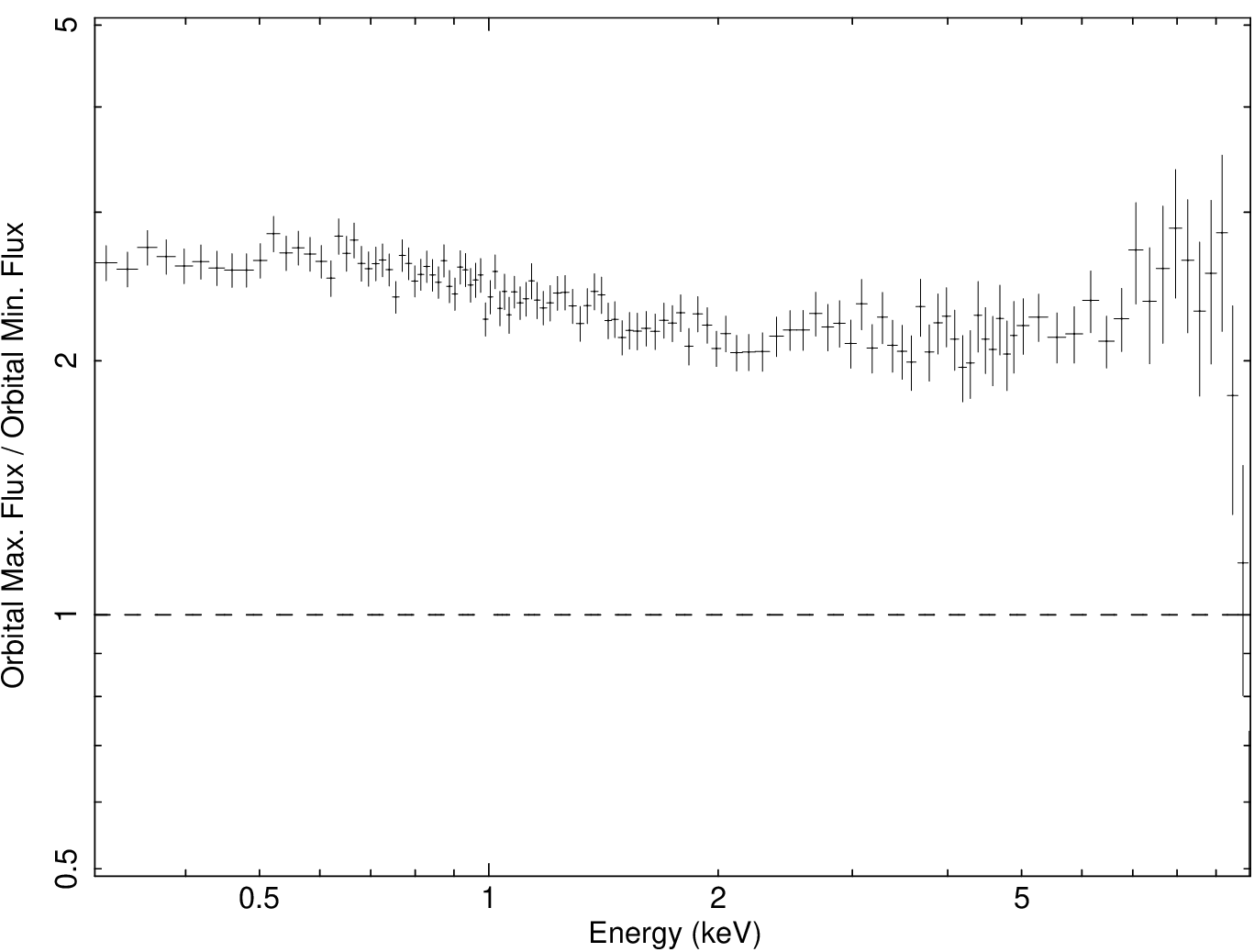}
\includegraphics[scale=0.35,angle=270]{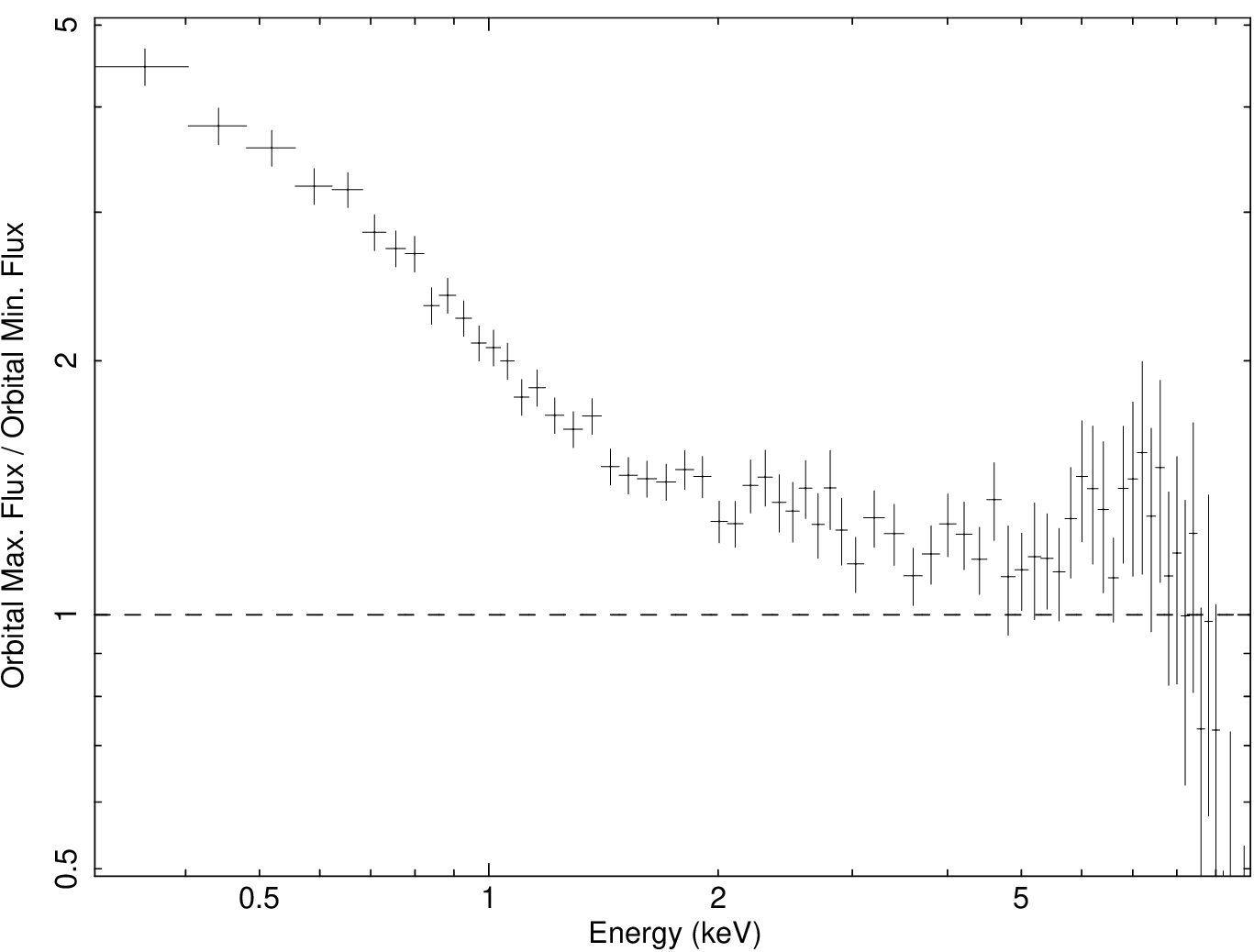}}
\caption{The left-hand panel shows the ratio of the count rate
spectrum of the orbital maximum to the spectrum of the orbital
minimum in the year 2000 data. The right-hand panel shows the ratio
of the count rate spectrum of the orbital maximum to the spectrum of
the orbital minimum in the year 2003 data.}
\end{figure*}

\newpage

\begin{figure*}
\centerline{
\includegraphics[width=7cm,height=13cm, angle=270]{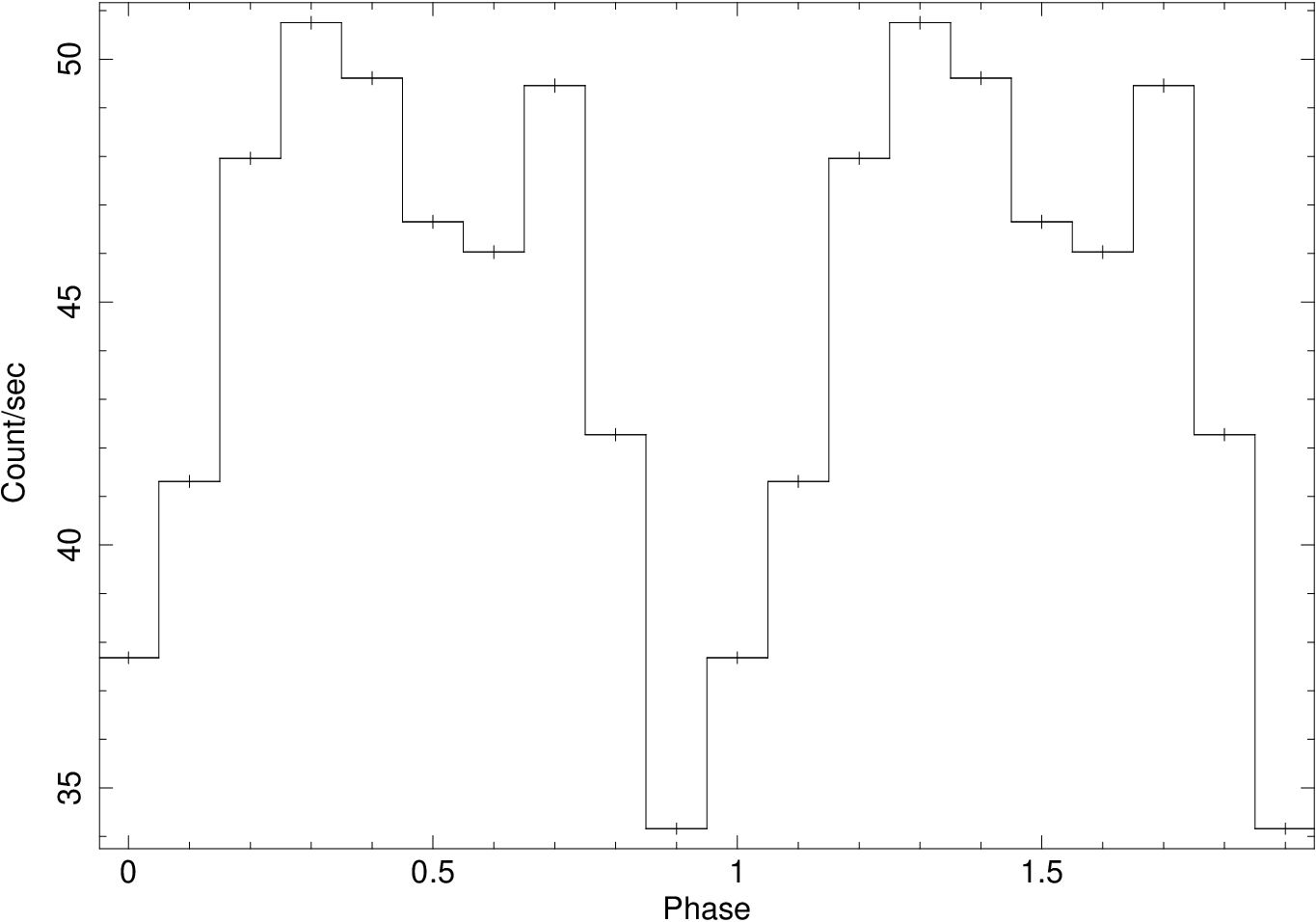}}
\centerline{
\includegraphics[width=7cm,height=13cm, angle=270]{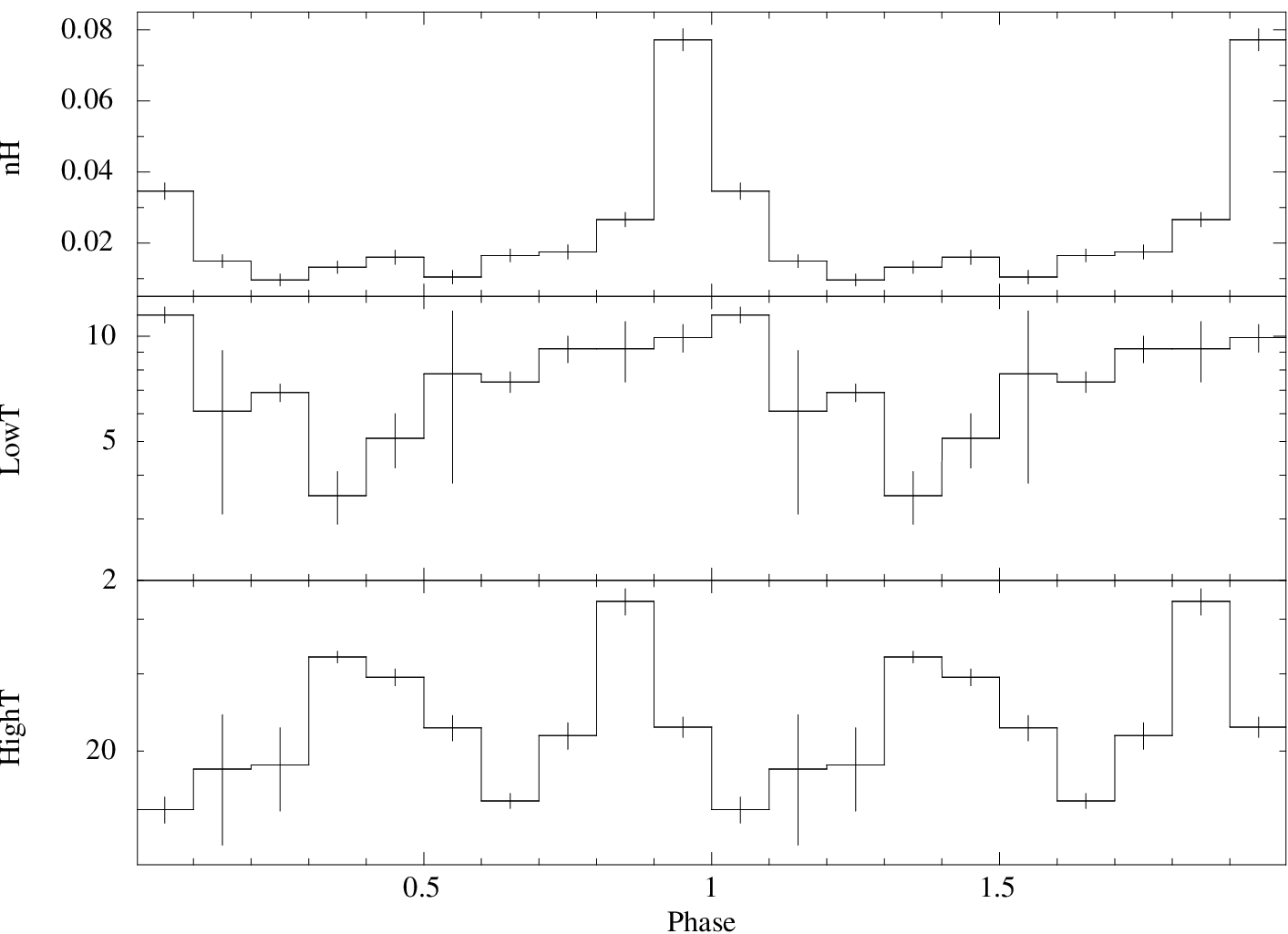}}
\centerline{
\includegraphics[width=7cm,height=13cm, angle=270]{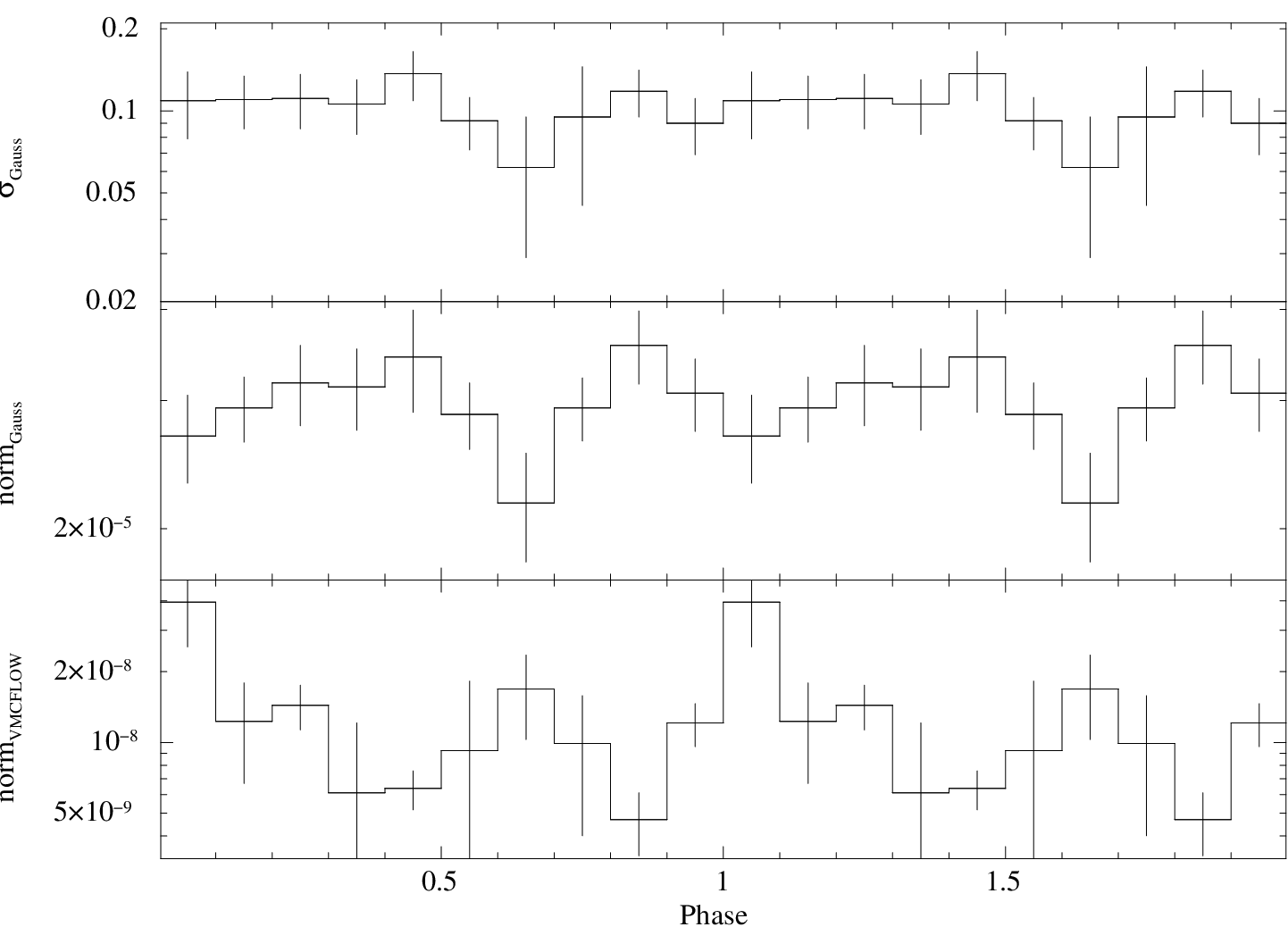}}

\caption{The light curve of the year 2000 data folded over the
orbital period of 98 minutes using the ephemeris T =
2437699.94179+0.068233846(4)E (top panel) and plots of the spectral
parameters derived from the orbital phase-resolved spectroscopy over
the same orbital phase range. The $\sigma$$_{\rm{Gauss}}$ and
norm$_{\rm{Gauss}}$ are parameters for the 6.7 keV line. The lower
boundary errors were used in plotting the $highT$ parameter. }
\end{figure*}

\newpage

\begin{figure*}
\centerline{
\includegraphics[scale=0.5,angle=270]{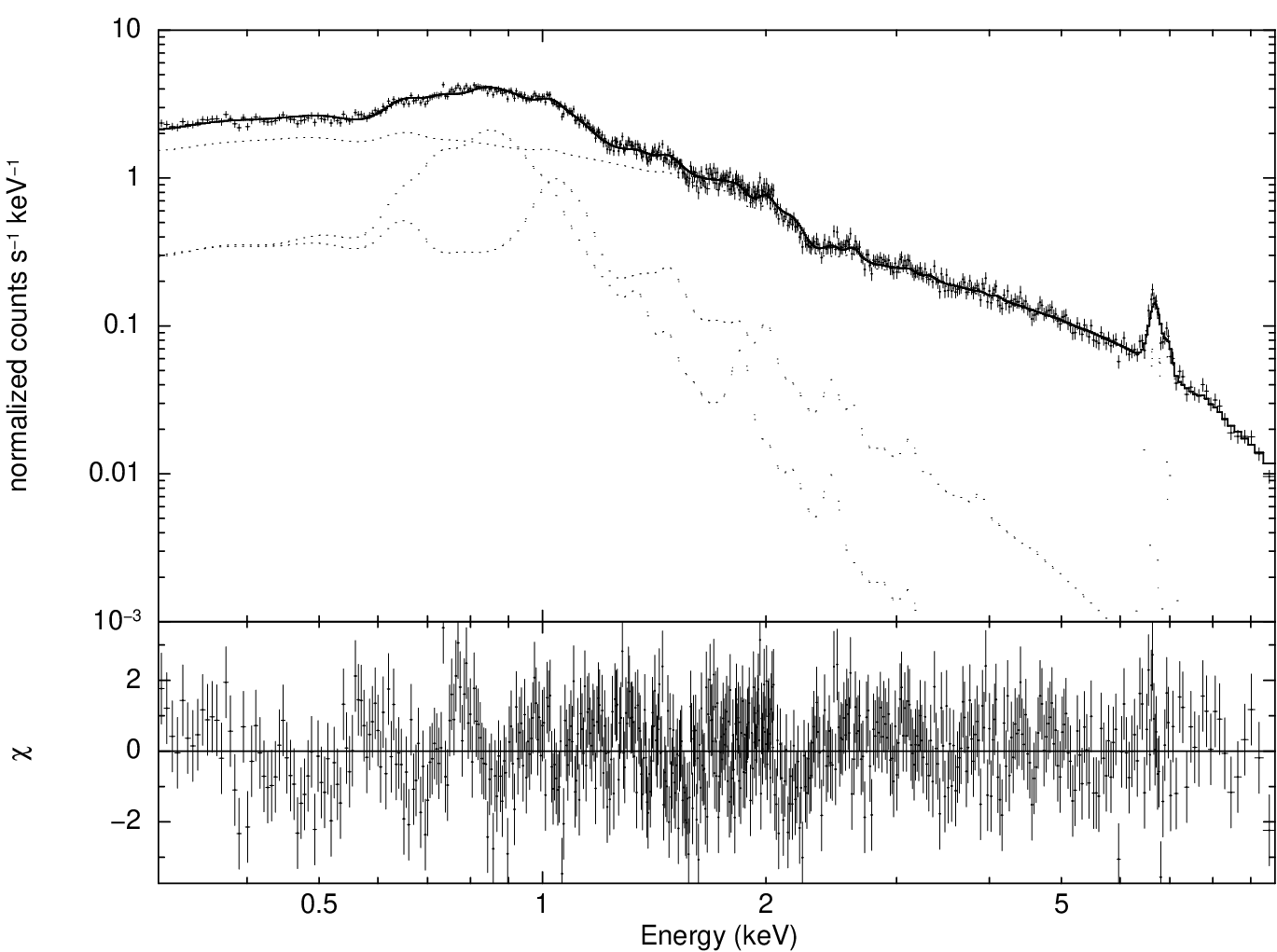}}
\centerline{
\includegraphics[scale=0.5,angle=270]{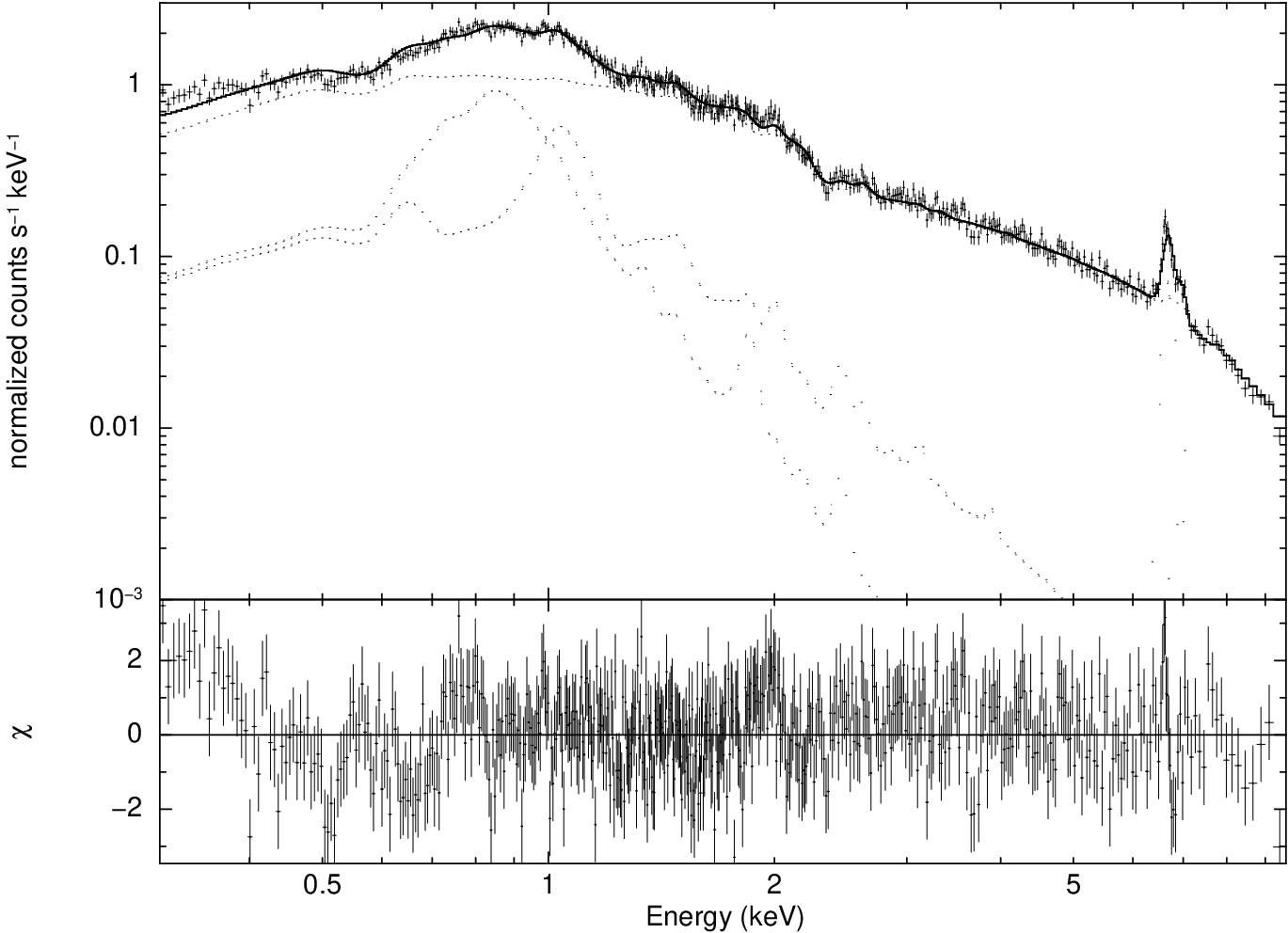}}
\caption{The top panel shows the composite model
(WABS*(MEKAL+MEKAL+GAUSS+GAUSS+VMCFLOW)) fitted to the spectrum of
orbital maximum (phase 0.3) in the year 2000 data.
 The bottom panel shows the same
composite model fitted to the spectrum of the orbital minimum (phase
0.9) in the year 2000 data. The crosses show the data with error
bars, solid lines show the composite model, the dotted lines show
the individual models and the panels underneath show the residuals
in standard deviations.}
\end{figure*}

\newpage

\begin{figure*}
\centerline{
\includegraphics[width=7cm,height=13cm,angle=270]{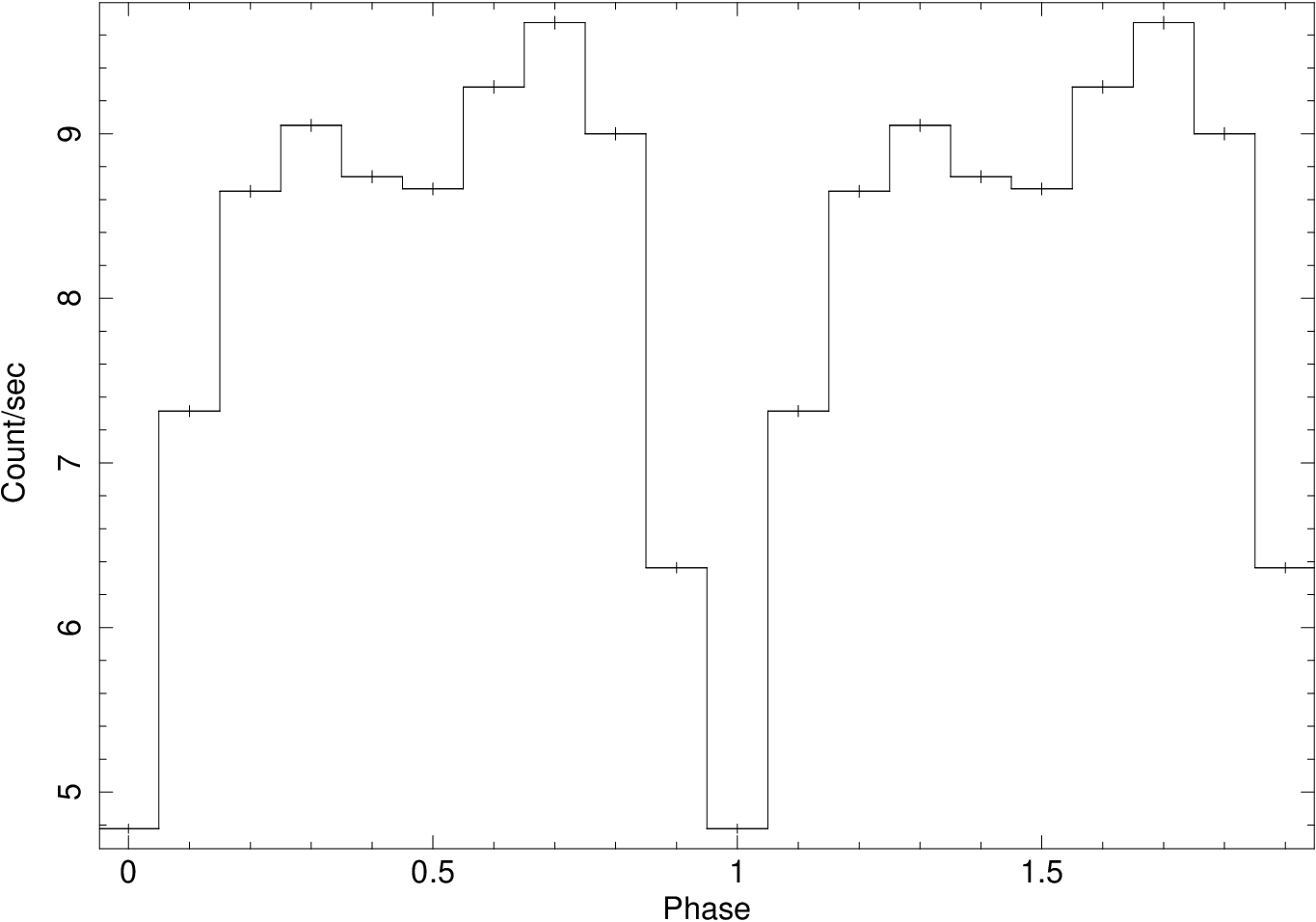}}
\centerline{
\includegraphics[width=7cm,height=13cm,angle=270]{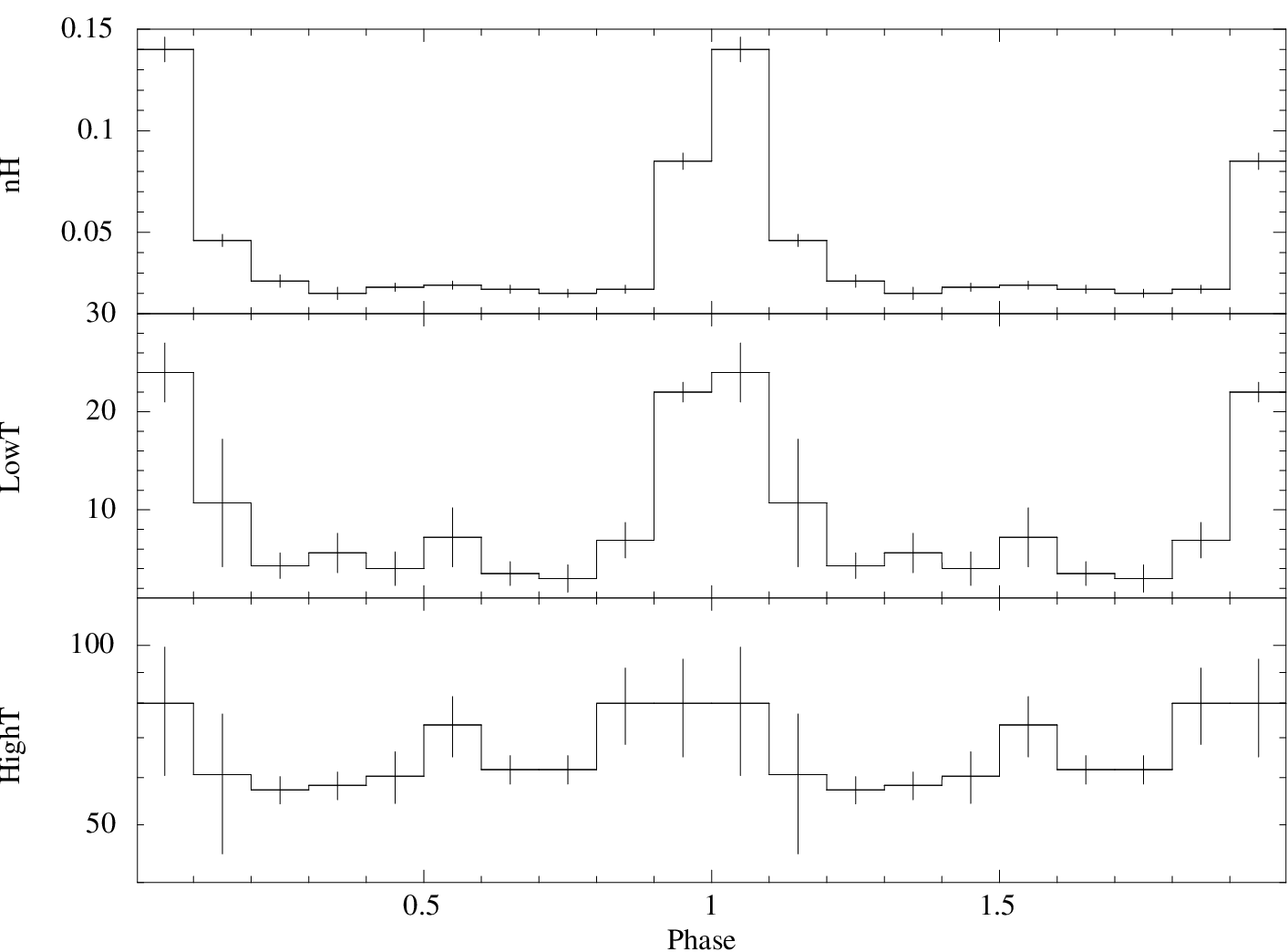}}
\centerline{
\includegraphics[width=7cm,height=13cm,angle=270]{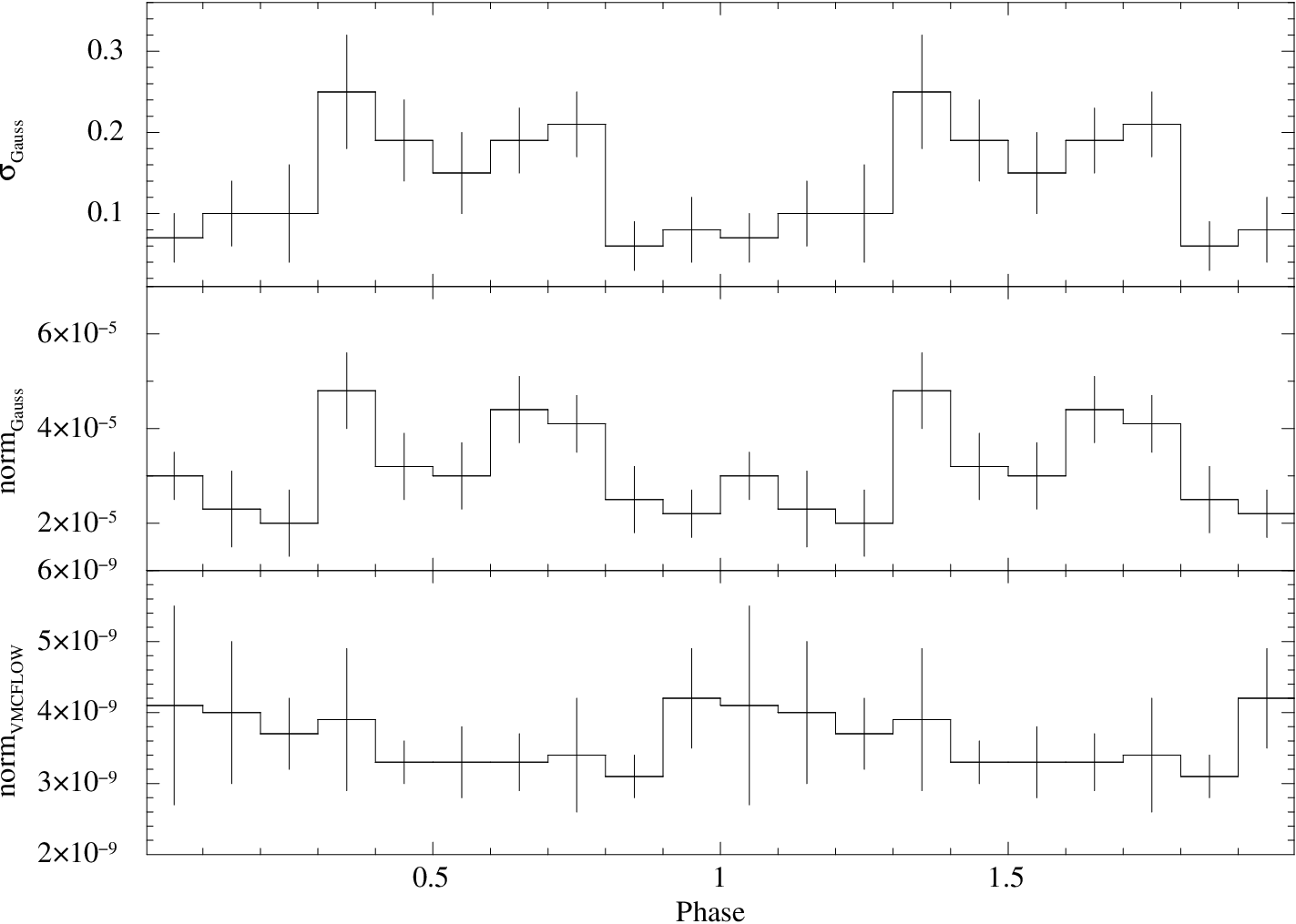}}

\caption{The light curve of the year 2003 data folded over the
orbital period of 98 minutes using the ephemeris T =
2437699.94179+0.068233846(4)E (top panel) and plots of the spectral
parameters derived from the orbital phase-resolved spectroscopy over
the same orbital phase range. The $\sigma$$_{\rm{Gauss}}$ and
norm$_{\rm{Gauss}}$ are parameters for the 6.7 keV line. The lower
boundary errors were used in plotting the $highT$ parameter.}

\end{figure*}

\newpage

\begin{figure*}
\centerline{
\includegraphics[scale=0.5,angle=270]{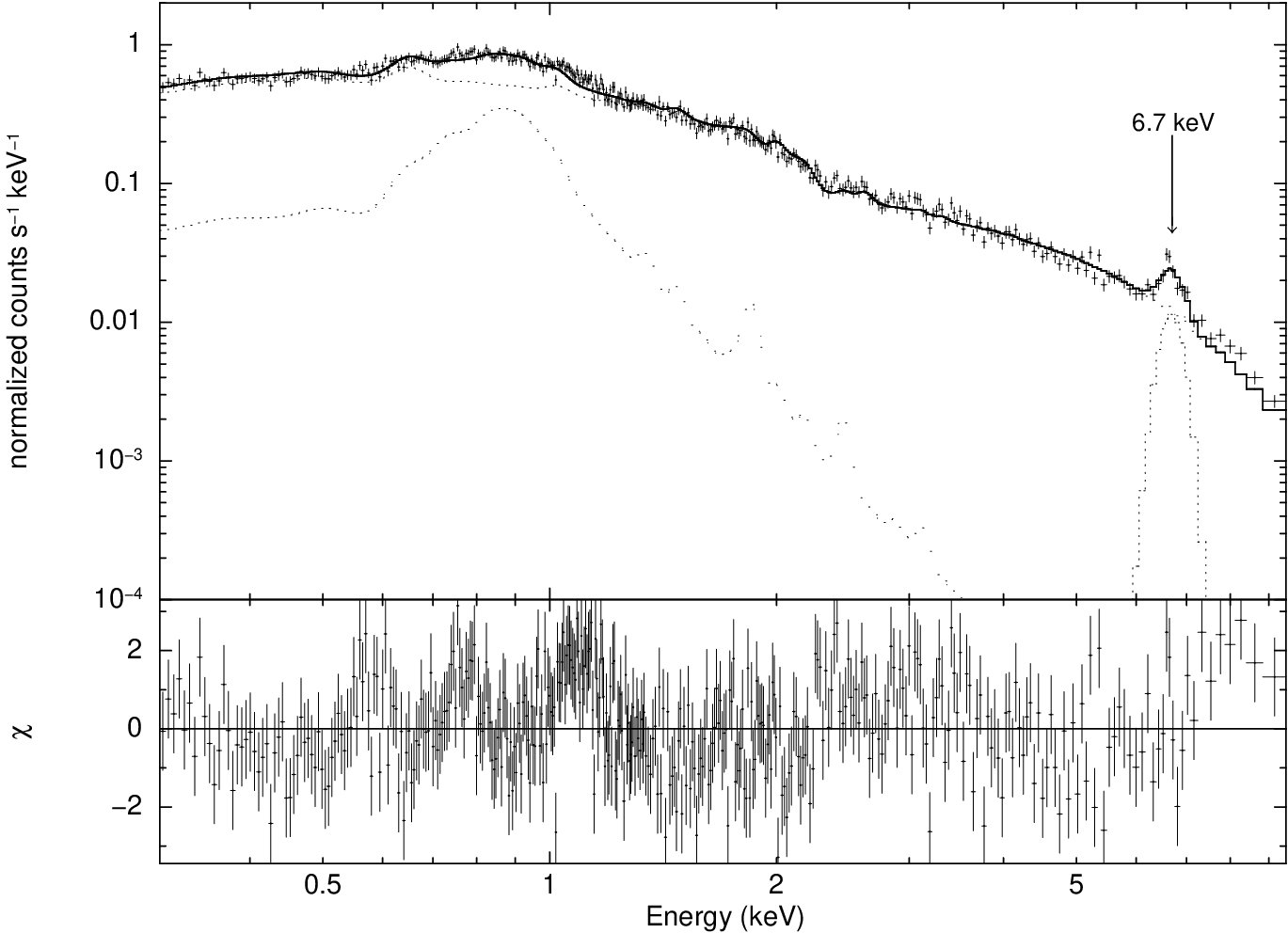}}
\centerline{
\includegraphics[scale=0.5,angle=270]{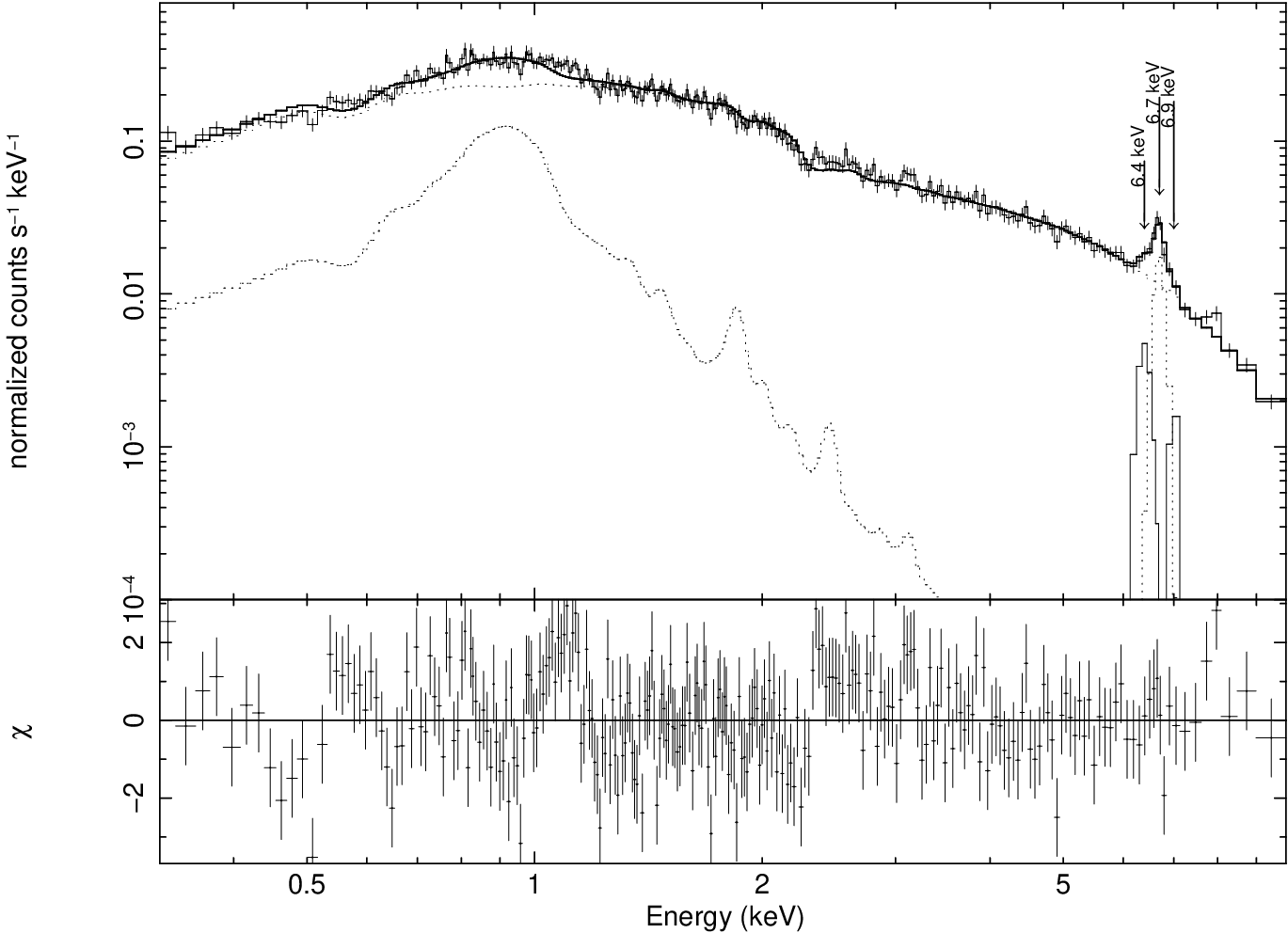}}
\caption{The top panel shows the composite model
(PCFABS*(MEKAL+MEKAL+GAUSS+VMCFLOW)) fitted to the spectrum of
orbital maximum (phase 0.4) in the year 2003 data. Note that only
6.7 keV line is present. The bottom panel shows the same composite
model fitted to the spectrum of the orbital minimum (phase 0) in the
year 2003 data. Note that 6.4 keV, 6.7 keV and 6.9 keV Fe emission
lines are present. The crosses show the data with error bars, solid
lines show the composite model, the dotted lines show the individual
models and the panels underneath show the residuals in standard
deviations.}
\end{figure*}

\newpage

\begin{figure*}
%\begin{center}
%\begin{array}{cc}
\centerline{
\includegraphics[scale=0.35,angle=270]{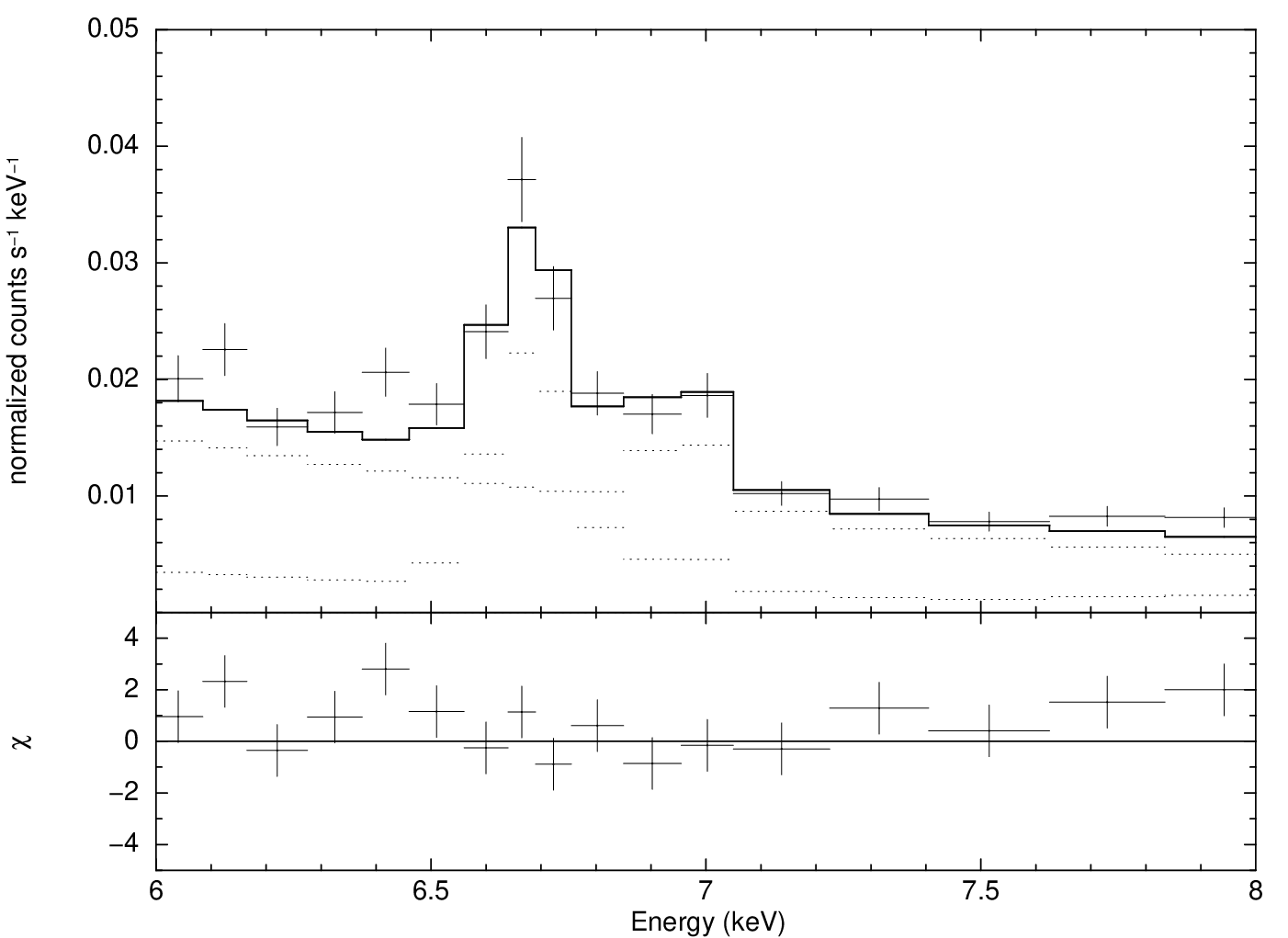}
\includegraphics[scale=0.35,angle=270]{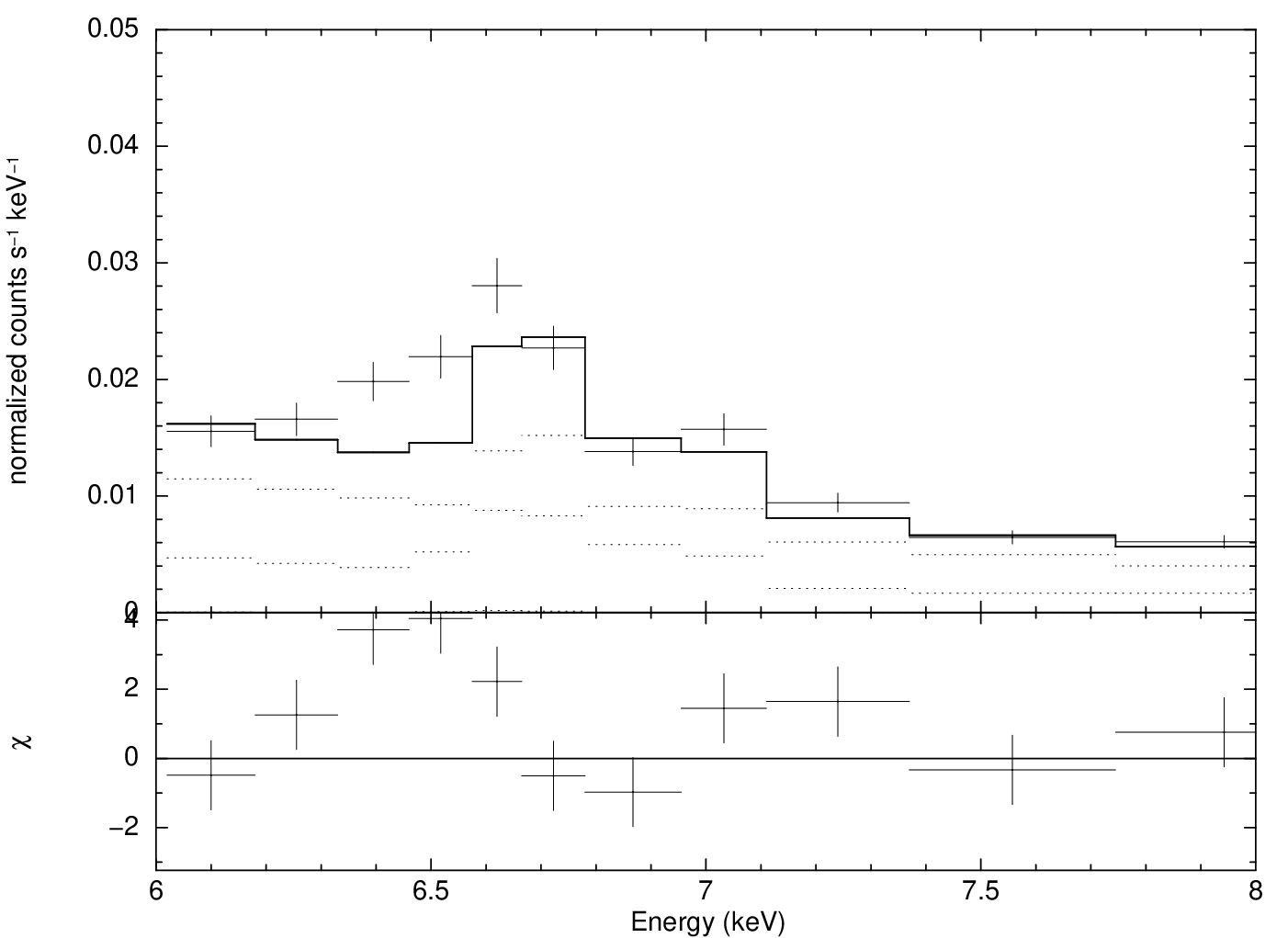}} 
\centerline{
\includegraphics[scale=0.35,angle=270]{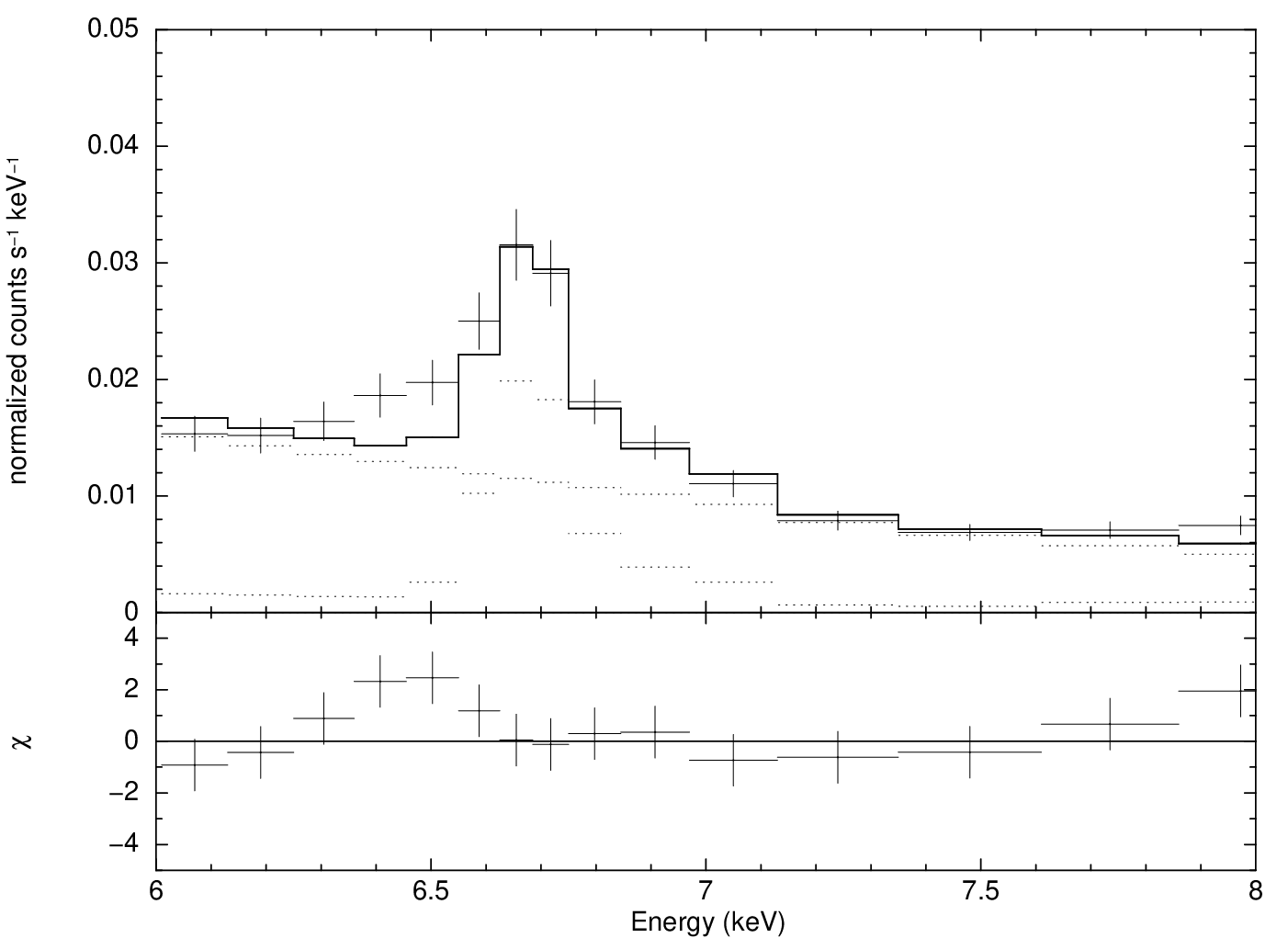}
\includegraphics[scale=0.35,angle=270]{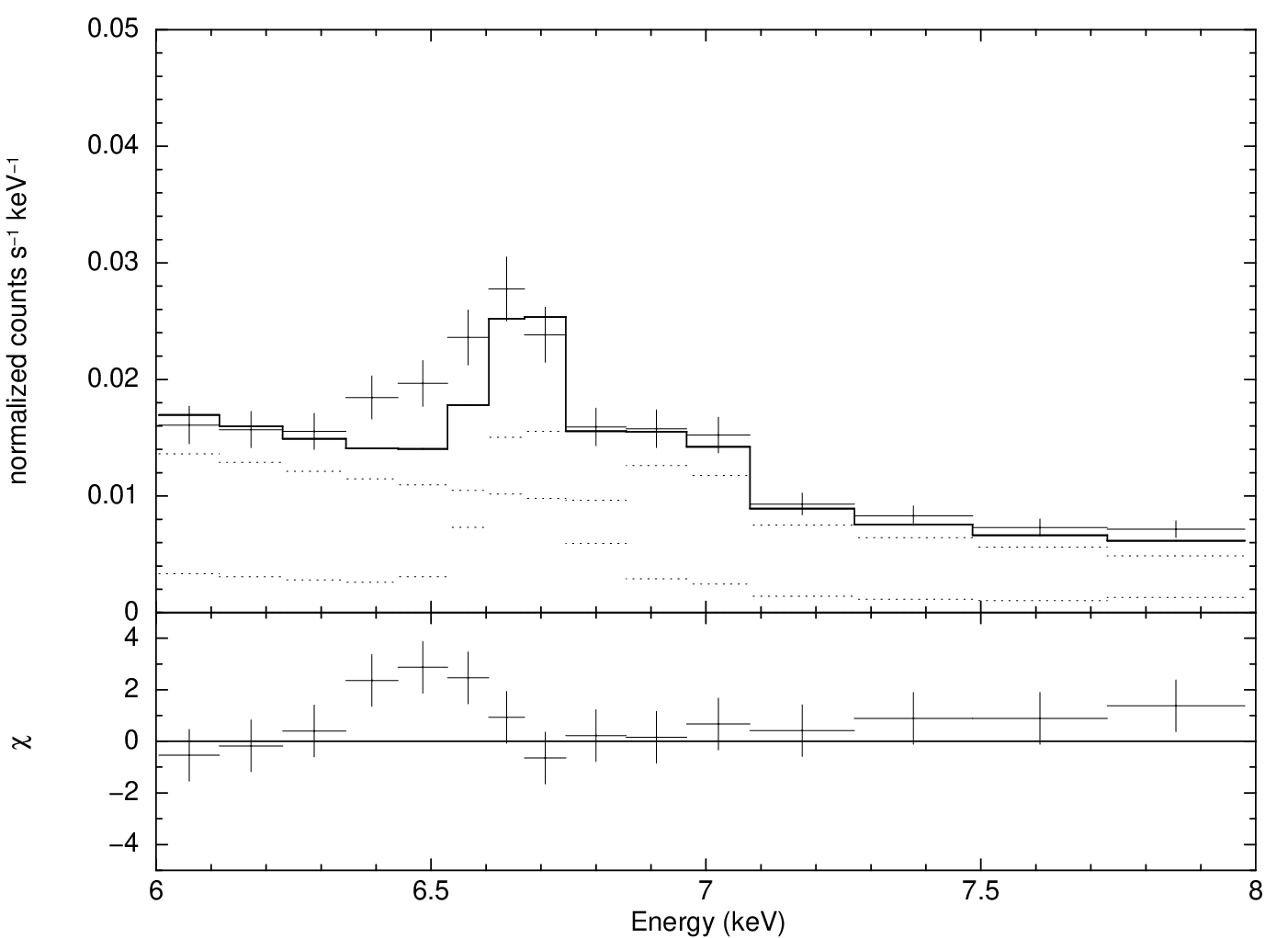}} 
\includegraphics[scale=0.35,angle=270]{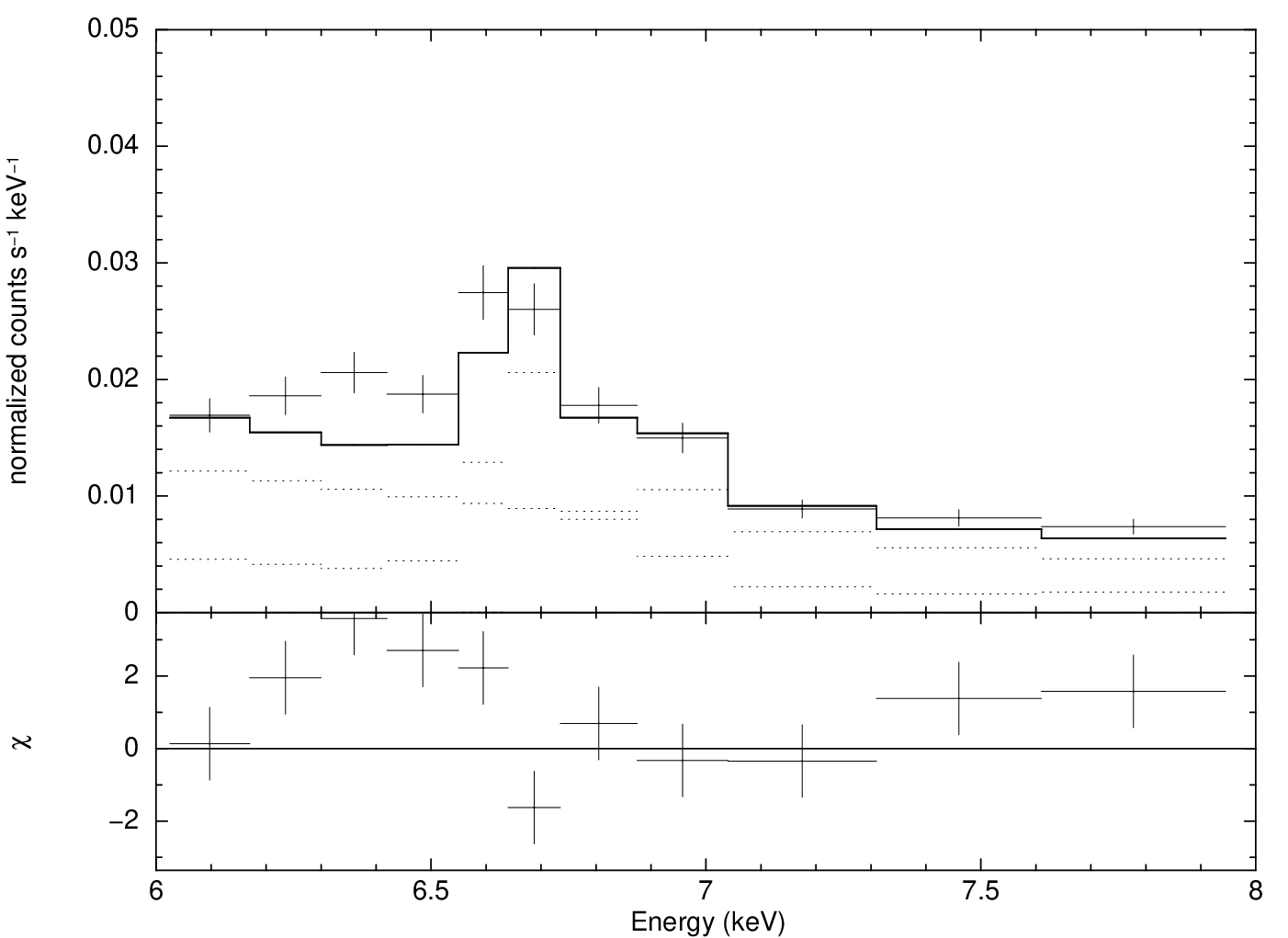} 

%\end{array}
%\end{center}
\caption{The orbital phase-resolved spectra of 2003 data between
phases 0.9 to 1.3 from left to right and down, respectively. The energy range is
between 6 and 8 keV to focus on the emission lines. The spectra were
fitted with 4 MEKAL models to illustrate the excess at 6.4 keV.  }
\end{figure*}

\newpage

\begin{figure*}
\centerline{
\includegraphics[scale=0.35,angle=270]{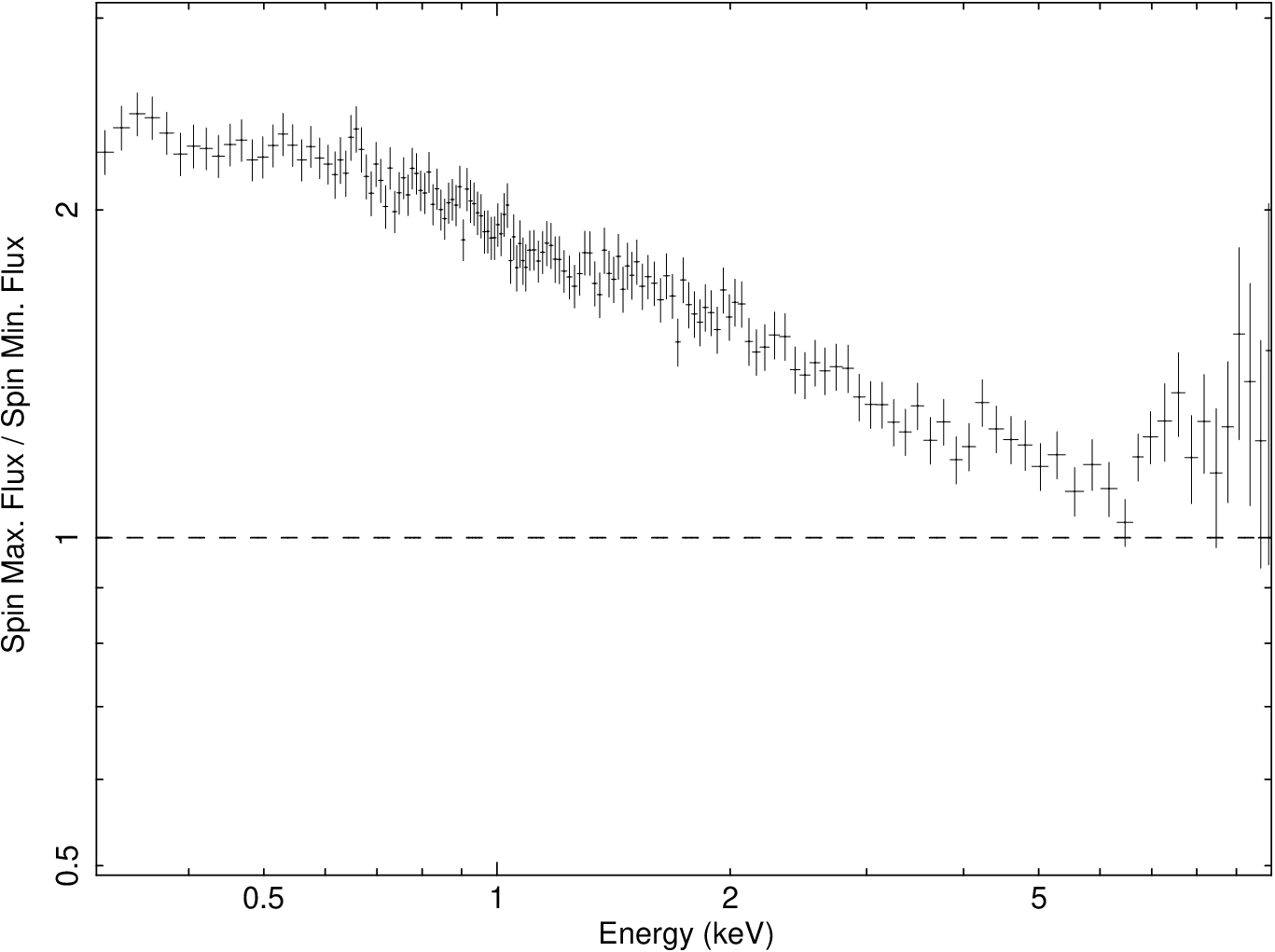}
\includegraphics[scale=0.35,angle=270]{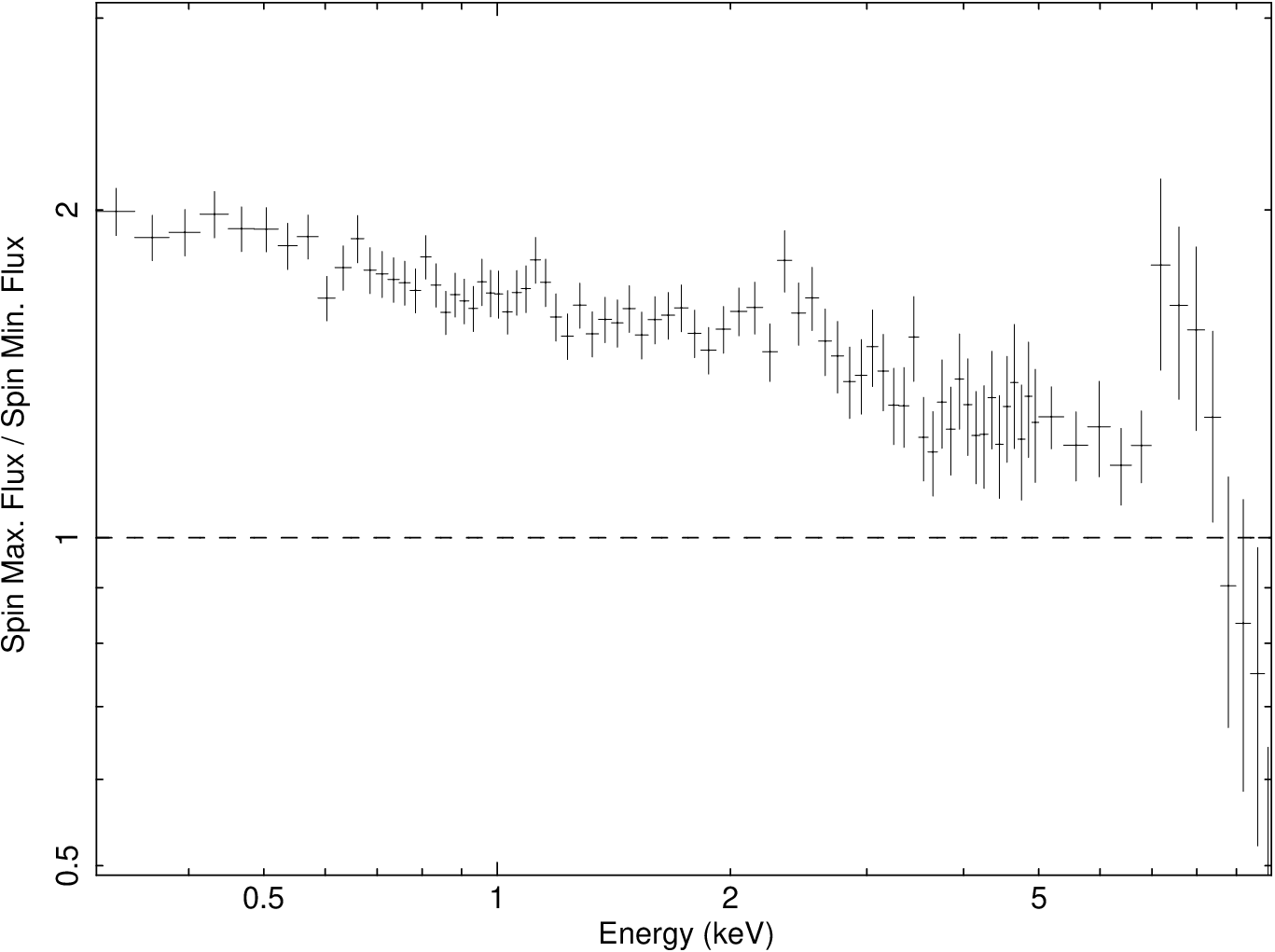}}
\caption{The left-hand panel shows the ratio of the count rate
spectrum of the spin maximum phase to the spectrum of the spin
minimum phase in the year 2000 data. The right-hand panel shows the
ratio of the count rate spectrum of the spin maximum to the spectrum
of the spin minimum in the year 2003 data.}
\end{figure*}

\newpage

\begin{figure*}
\centerline{
\includegraphics[scale=0.5,angle=270]{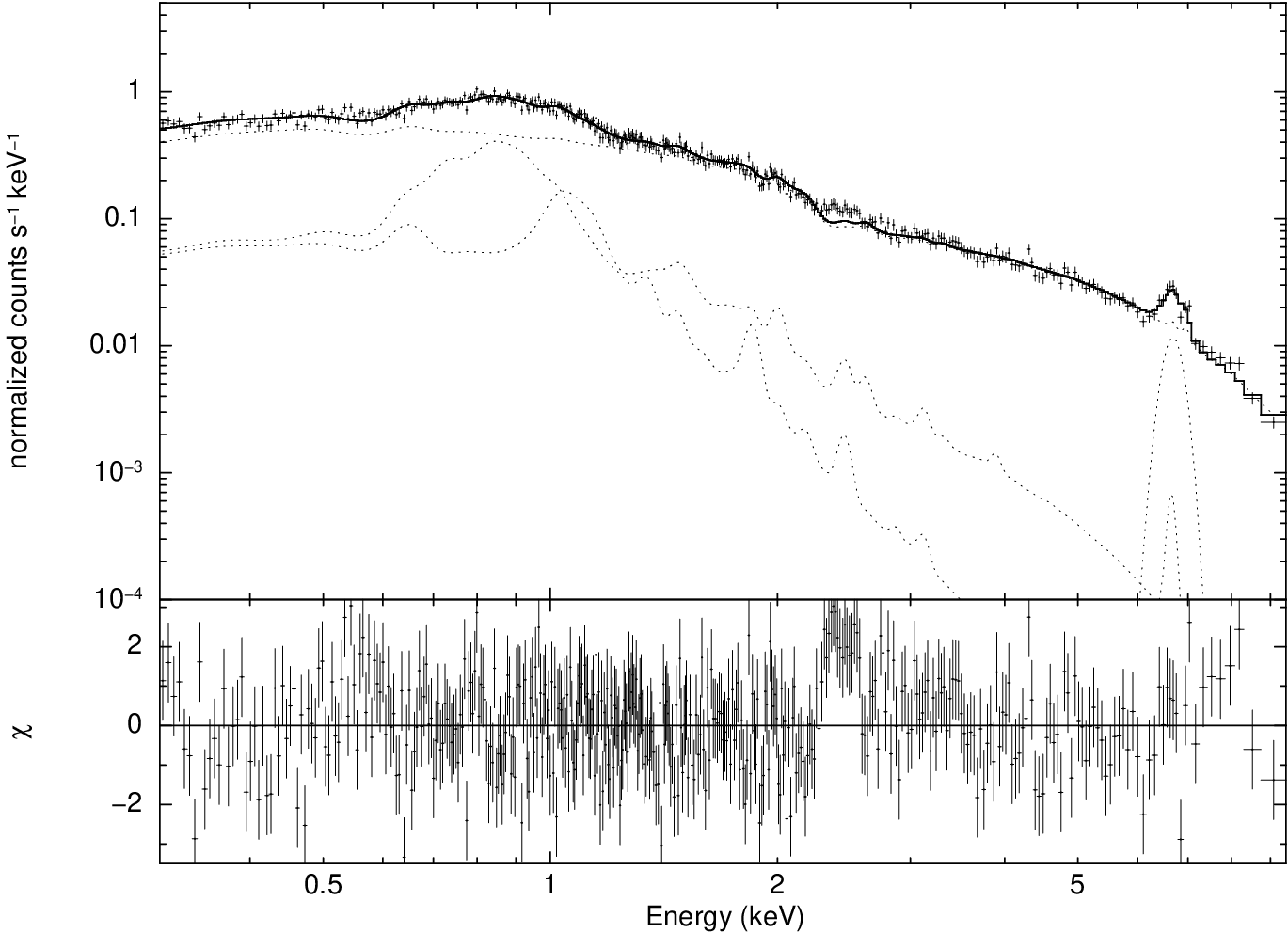}}
\centerline{
\includegraphics[scale=0.5,angle=270]{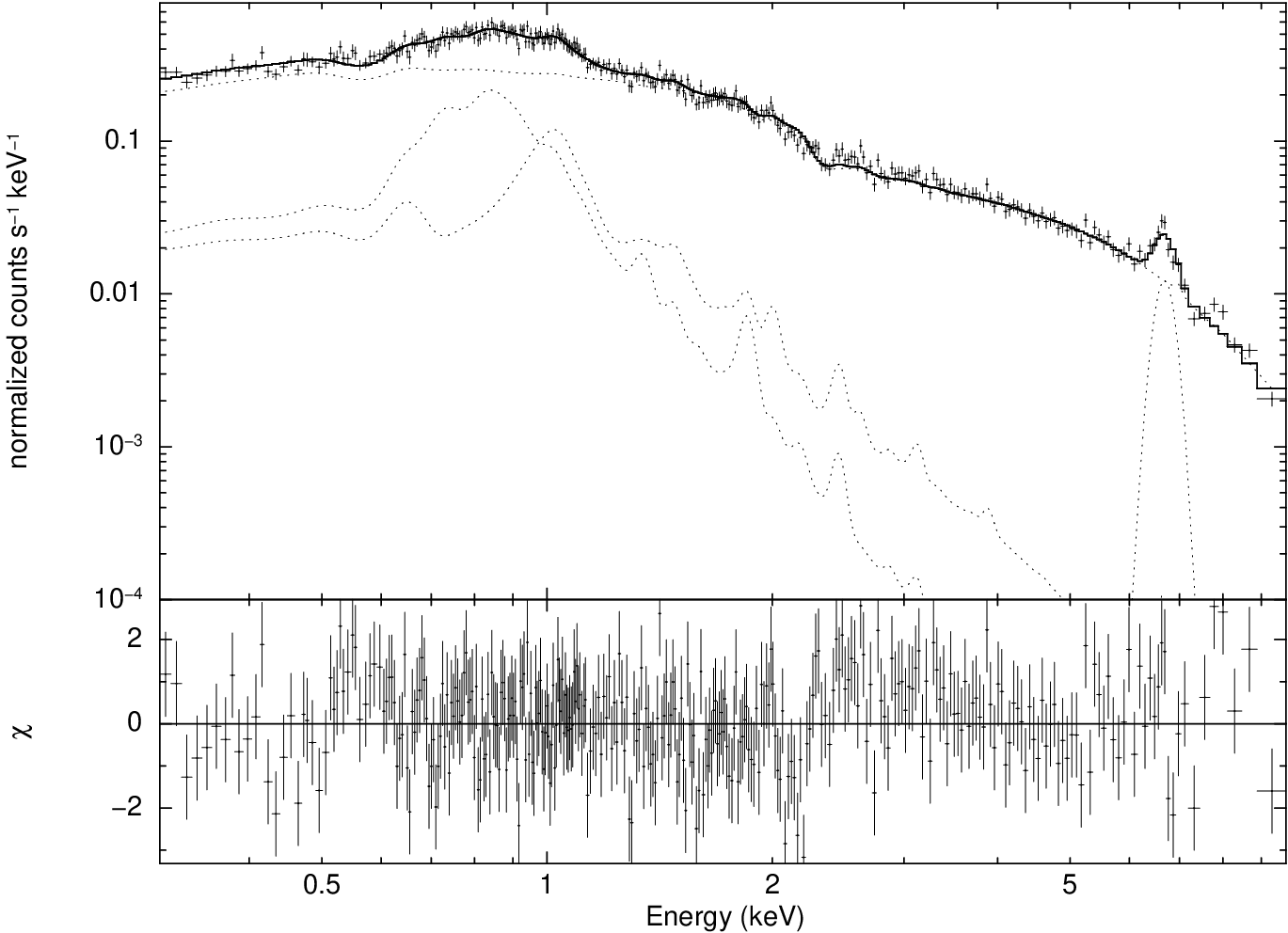}}
\caption{The top panel displays the composite model
(PCFABS*(MEKAL+MEKAL+GAUSS+VMCFLOW)) fitted to the spectrum of the
spin maximum (phase 0.7) in the year 2003 data. The bottom panel
displays the same composite model fitted to the spectrum of the spin
minimum (phase 0.3) in the year 2003 data. The crosses show the data
with error bars, solid lines show the composite model, the dotted
lines show the individual models and the panels underneath show the
residuals in standard deviations.}
\end{figure*}

\newpage

\begin{figure*}
\centerline{
\includegraphics[width=7cm,height=13cm,angle=270]{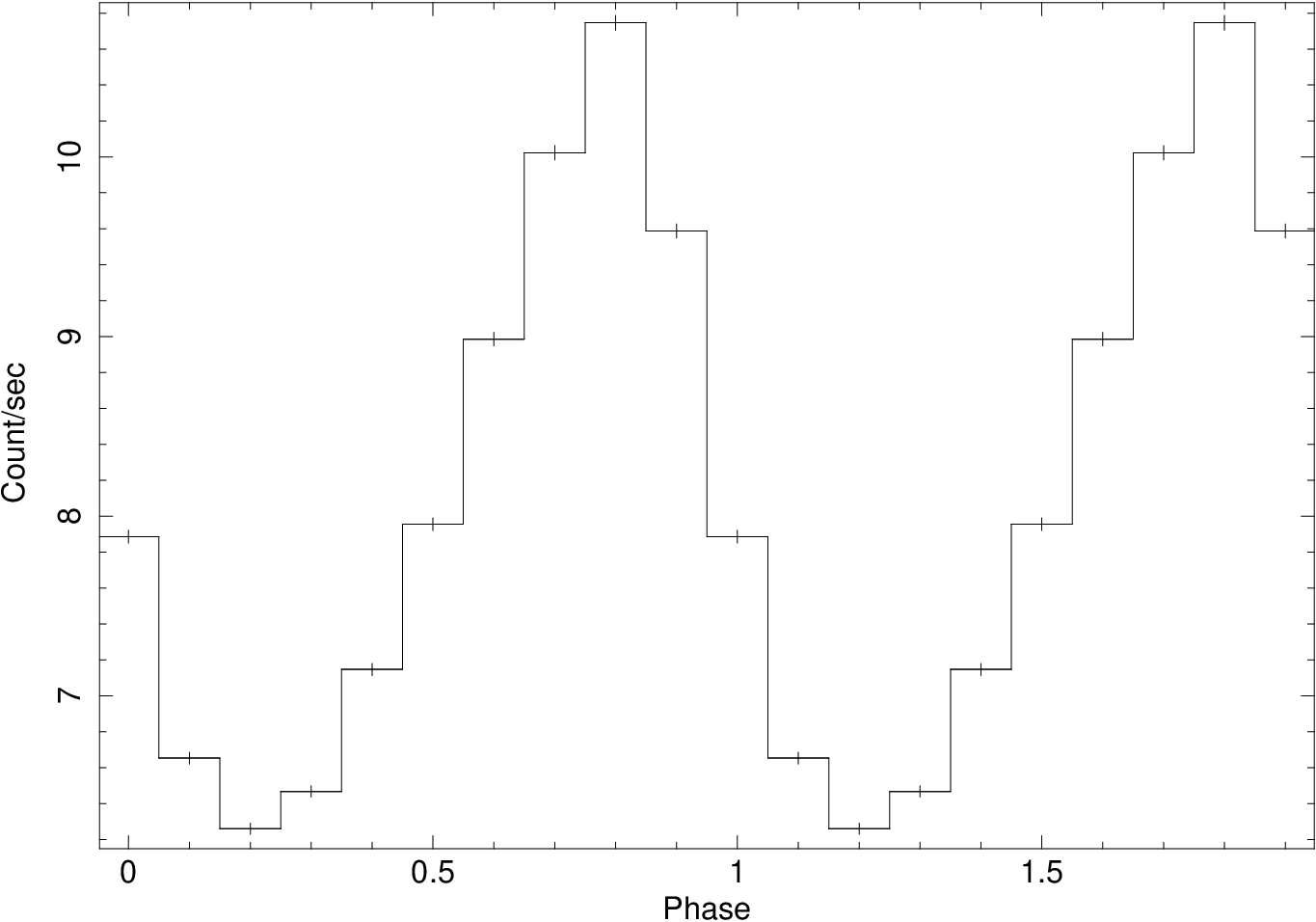}}
\centerline{
\includegraphics[width=7cm,height=13cm,angle=270]{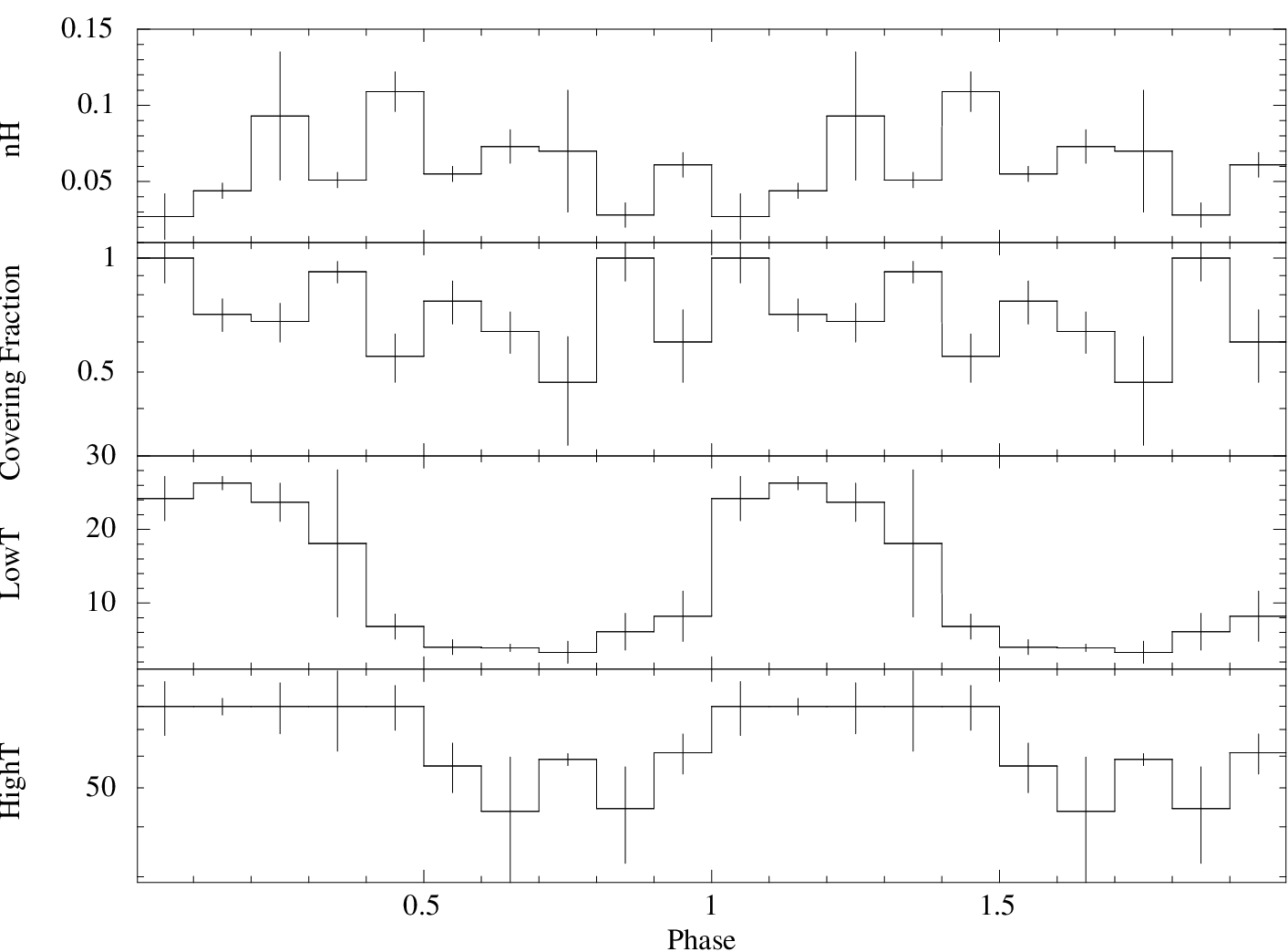}}
\centerline{
\includegraphics[width=7cm,height=13cm,angle=270]{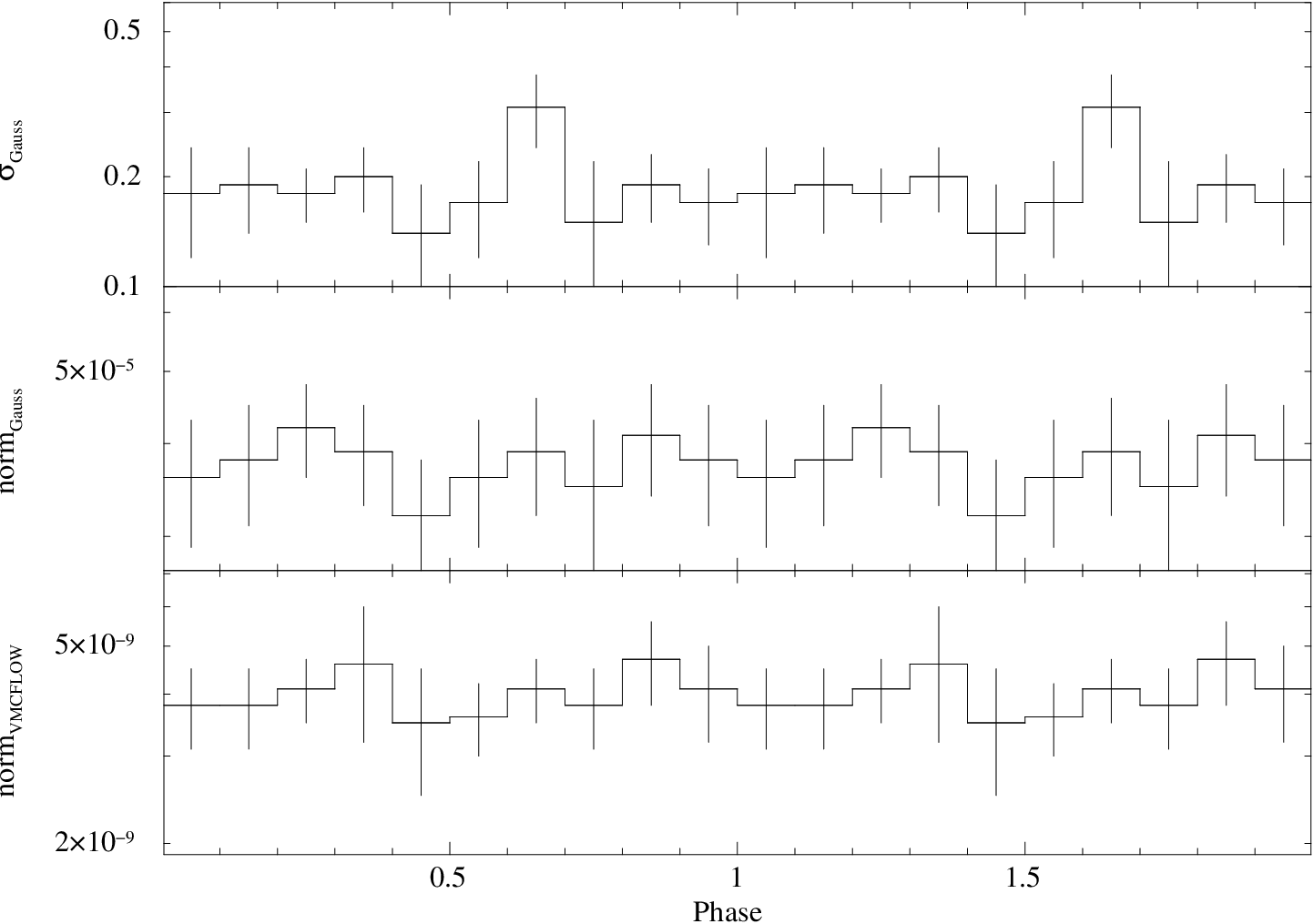}}

\caption{The light curve of the year 2003 data folded over the spin
period of 67 minutes using the ephemeris T = 2437699.8914(5) +
0.046546504(9)E (top panel) and plots of the spectral parameters
derived from the spin phase-resolved spectroscopy over the same spin
phase range. The $\sigma$$_{\rm{Gauss}}$ and $norm_{\rm{Gauss}}$ are
parameters for the 6.7 keV line. The lower boundary errors were used
in plotting the $highT$ parameter.}
\end{figure*}

\newpage

\begin{figure*}
\centerline{
\includegraphics[scale=0.5,angle=270]{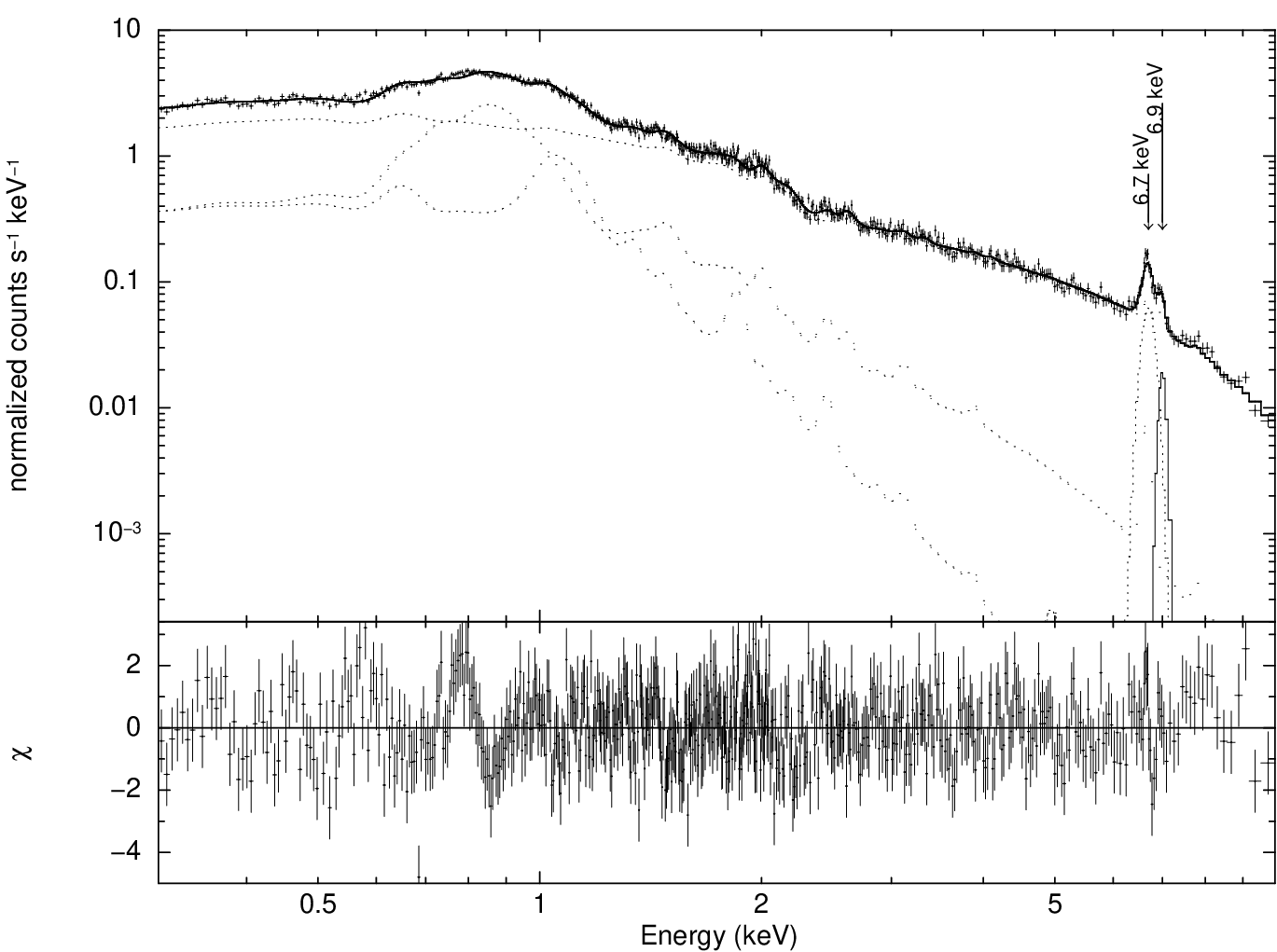}}
\centerline{
\includegraphics[scale=0.5,angle=270]{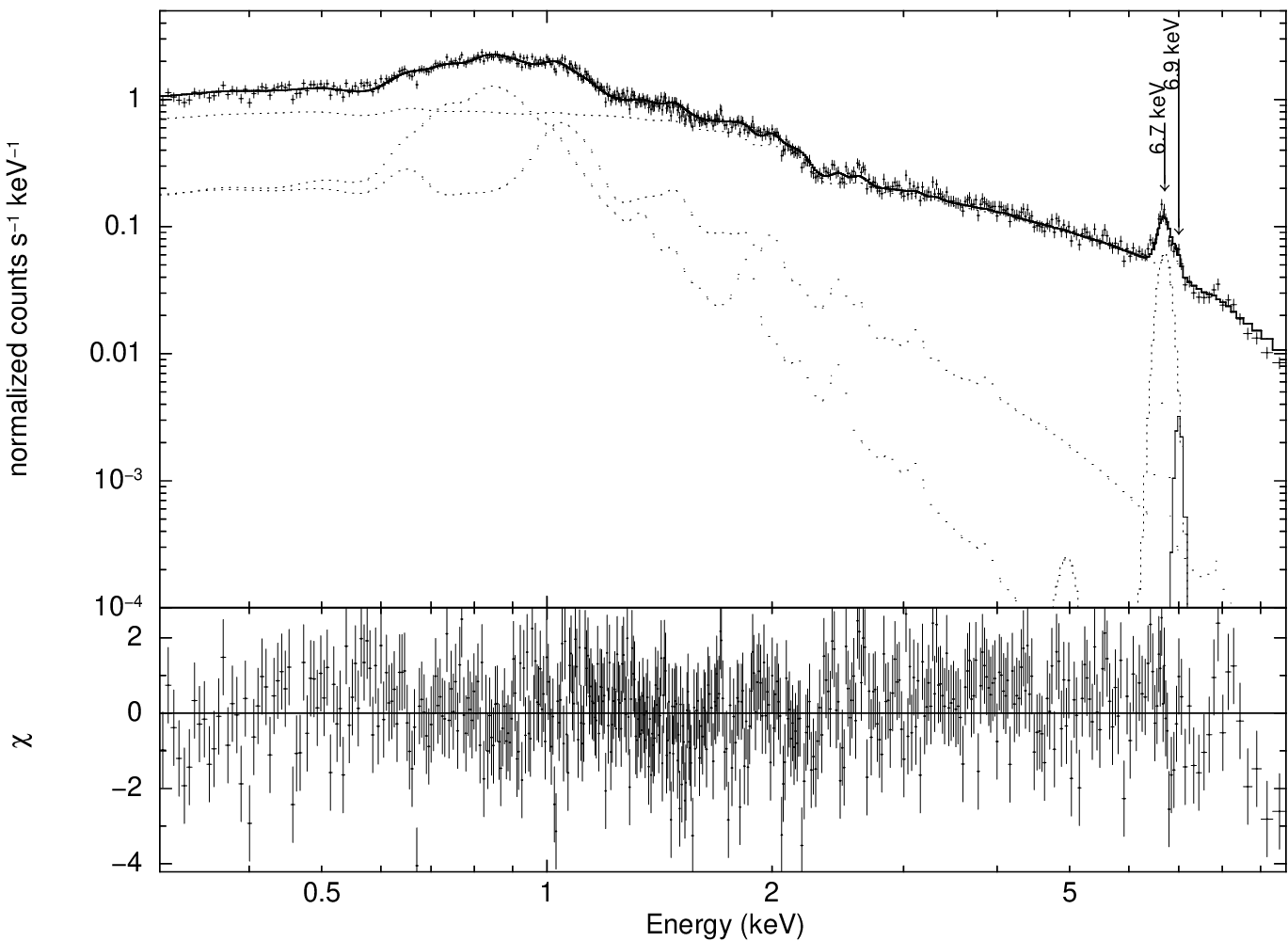}}
\caption{The top panel displays the composite model
(PCFABS*(MEKAL+MEKAL+GAUSS+GAUSS+VMCFLOW)) fitted to the spectrum of
the spin maximum (phase 0.9) in the year 2000 data. The 6.7 keV and
6.9 keV emission lines are explicitly shown. The bottom panel
displays the same composite model fitted to the spectrum of the spin
minimum (phase 0.5) in the year 2000 data. The 6.7 keV and 6.9 keV
emission lines are explicitly shown. The crosses show the data with
error bars, solid lines show the composite model, the dotted lines
show the individual models and the panels underneath show the
residuals in standard deviations.}
\end{figure*}

\newpage

\begin{figure*}
\centerline{
\includegraphics[width=7cm,height=13cm,angle=270]{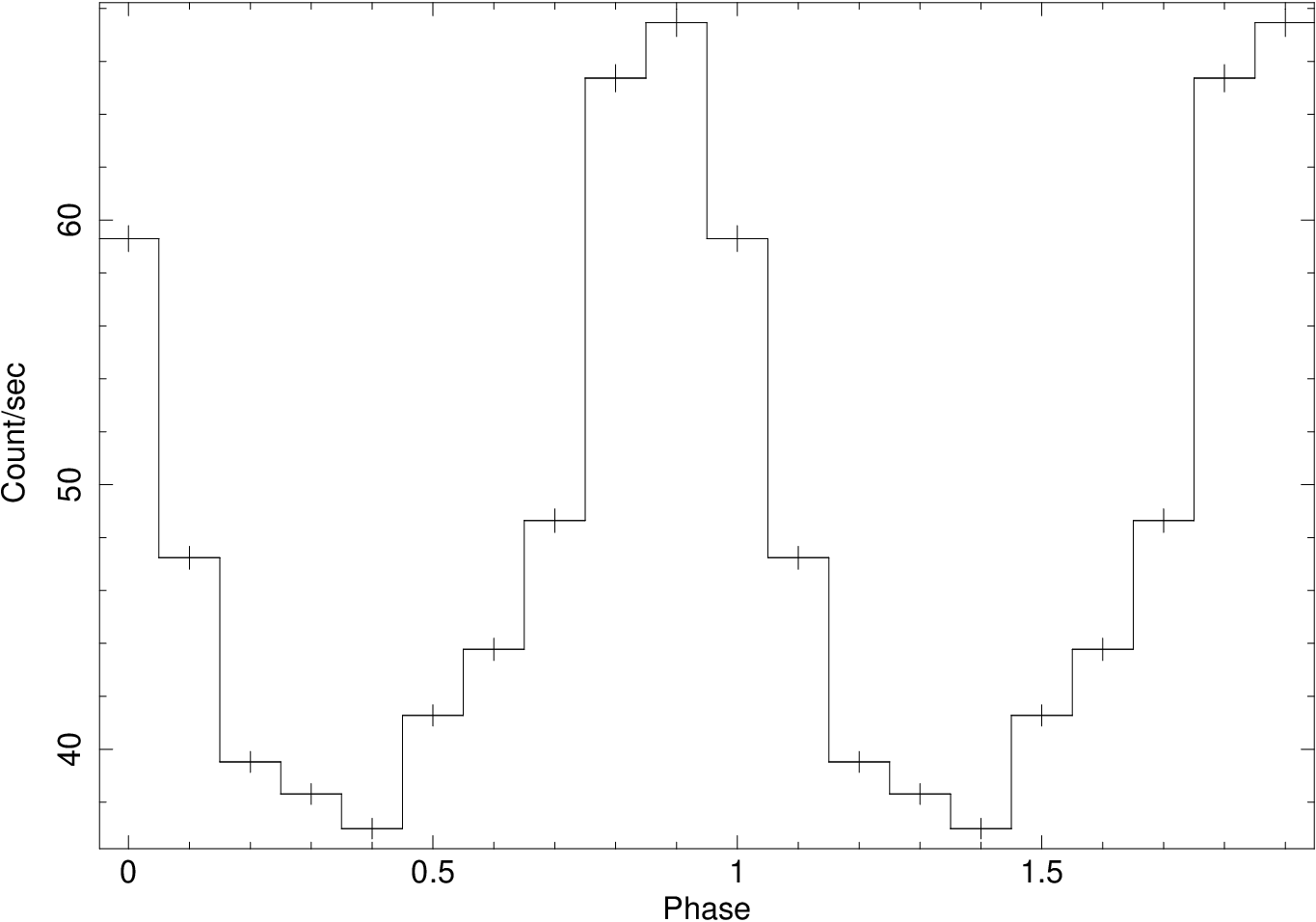}}
\centerline{
\includegraphics[width=7cm,height=13cm,angle=270]{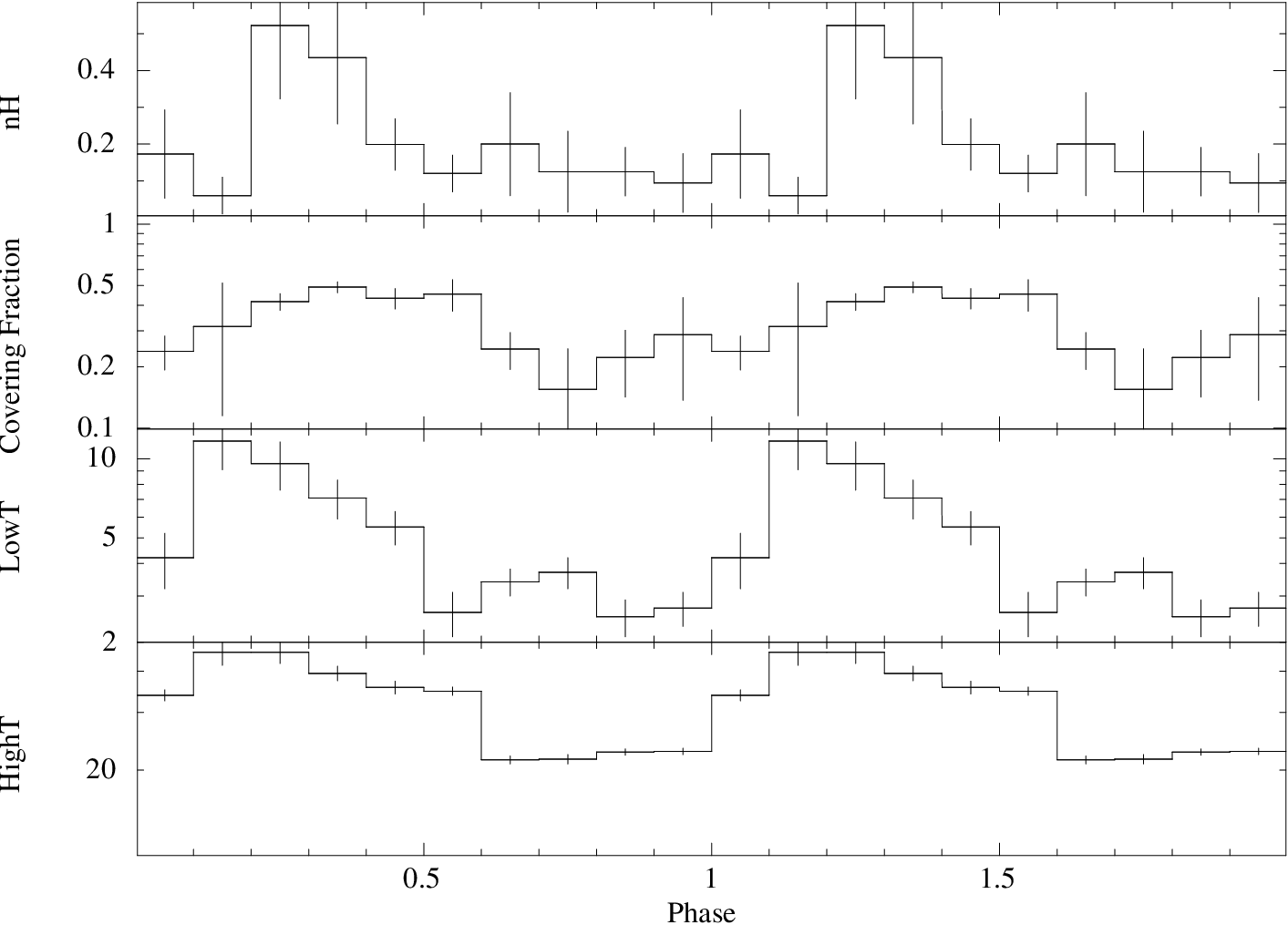}}
\centerline{
\includegraphics[width=7cm,height=13cm,angle=270]{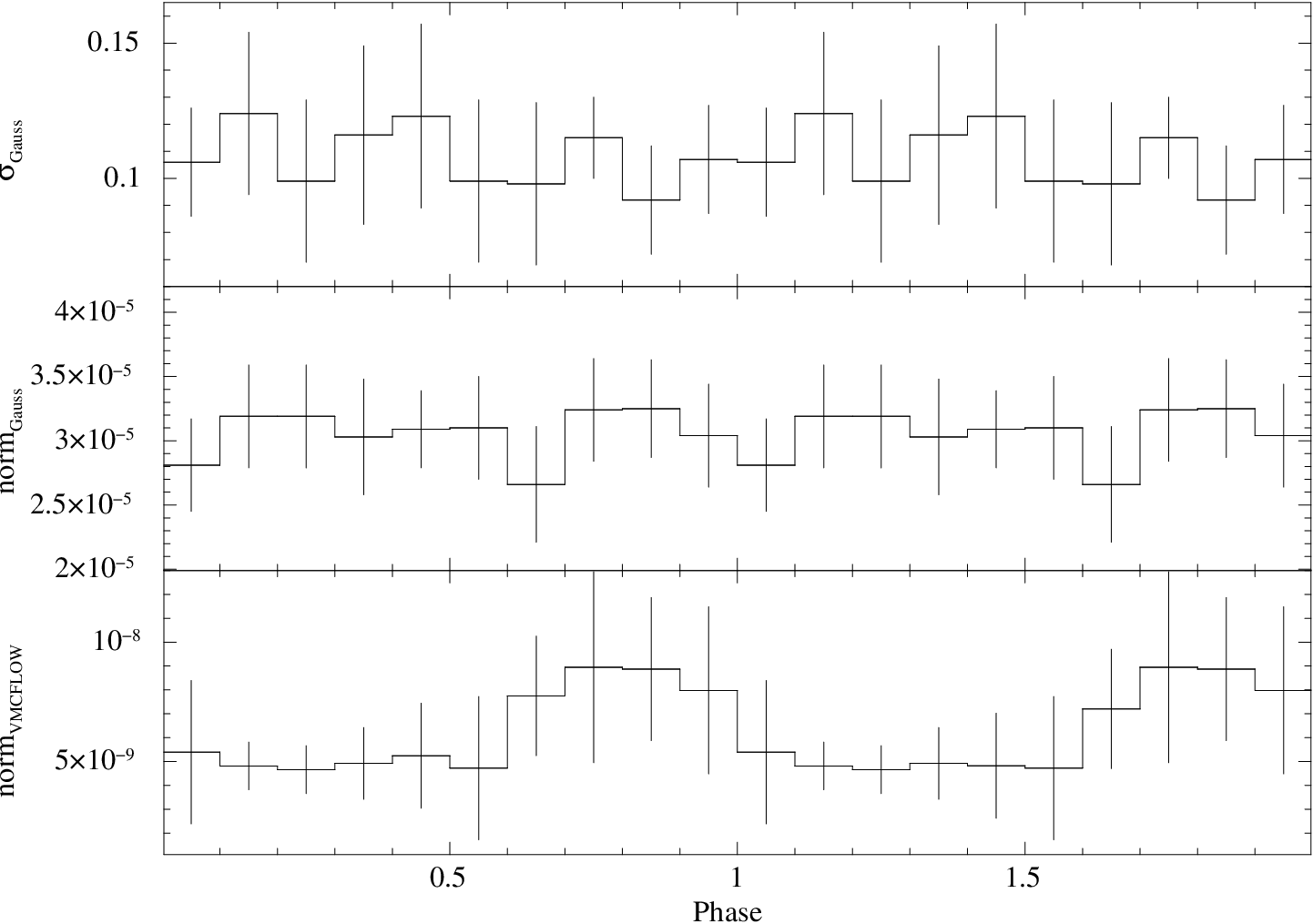}}

\caption{The light curve of the year 2000 data folded over the spin
period of 67 minutes using the ephemeris T = 2437699.8914(5) +
0.046546504(9)E (top panel) and plots of the spectral parameters
derived from the spin phase-resolved spectroscopy over the same spin
phase range. The $\sigma$$_{\rm{Gauss}}$ and $norm_{\rm{Gauss}}$ are
parameters for the 6.7 keV line. The lower boundary errors were used
in plotting the $highT$ parameter.}
\end{figure*}

\label{lastpage}

\end{document}